\documentclass[11pt, twocolumn ,trackchanges]{aastex631}
\usepackage{xcolor}
\usepackage{bm}
\usepackage{soul}
\definecolor{victoria}{rgb}{0.76, 0.36, 0.42}

\usepackage{float}

\usepackage{pgfplots}
\usepackage{multirow}
\pgfplotsset{compat=1.15}
\newcommand\bluesout{\bgroup\markoverwith{\textcolor{blue}{\rule[0.5ex]{2pt}{0.4pt}}}\ULon}

\date{\today}

\begin{document}

\title{Anisotropy of Density Fluctuations in the Solar Wind at 1 au}
\author[0009-0008-8723-610X]{Jiaming Wang}
\affiliation{Department of Physics and Astronomy, University of Delaware}
\author[0000-0002-7174-6948]{Rohit Chhiber}
\affiliation{Department of Physics and Astronomy, University of Delaware}
\affiliation{Heliophysics Science Division, NASA Goddard Space Flight Center}
\author[0000-0003-3891-5495]{Sohom Roy}
\affiliation{Department of Physics and Astronomy, University of Delaware}
\author[0000-0002-7341-2992]{Manuel E. Cuesta}
\affiliation{Department of Astrophysical Sciences, Princeton University}
\author[0000-0003-4168-590X]{Francesco Pecora}
\affiliation{Department of Physics and Astronomy, University of Delaware}
\author[0000-0003-2965-7906]{Yan Yang}
\affiliation{Department of Physics and Astronomy, University of Delaware}
\author[0000-0002-4305-6624]{Xiangrong Fu}
\affiliation{New Mexico Consortium, Los Alamos, NM 87544}
\affiliation{Los Alamos National Laboratory, Los Alamos, NM 87545}
\author[0000-0003-3556-6568]{Hui Li}
\affiliation{Los Alamos National Laboratory, Los Alamos, NM 87545}
\author[0000-0001-7224-6024]{William H. Matthaeus}
\affiliation{Department of Physics and Astronomy, University of Delaware}

% \maketitle

\begin{abstract}
    
A well-known property of solar
wind plasma turbulence is the observed anisotropy of the autocorrelations, or 
equivalently the spectra, of velocity and magnetic field fluctuations. Here we explore 
the related but apparently not well-studied issue of the anisotropy of plasma density fluctuations in the energy-containing and inertial 
ranges of solar wind turbulence. Using 10 years (1998-2008) of {\it in situ} data from the Advanced Composition Explorer (ACE) mission, we find that for all but the fastest wind category, the density correlation scale is slightly larger in 
directions quasi-parallel to the large-scale mean magnetic field
as compared to quasi-perpendicular directions. The correlation scale in fast wind is consistent with isotropic. The anisotropy as a function of the level of correlation is also explored. We find at small correlation levels,
i.e., at energy-containing scales and larger, the density fluctuations 
are close to isotropy for fast wind, and 
slightly favor more rapid decorrelation in perpendicular directions for slow and medium winds. At 
relatively smaller (inertial range) scales where the correlation values are larger, the sense of anisotropy is reversed in all speed ranges, implying a more ``slab-like'' structure, especially prominent in the fast wind samples.  We contrast this finding with published results on velocity and magnetic field correlations. 

\end{abstract}

\section{Introduction}

There are two well-known reasons for turbulent fluctuations in the solar wind to exhibit departures from statistical isotropy
\citep{BatchelorTHT,OughtonEA15}.
The first is solar wind expansion, which 
in the simplest terms imposes the radial coordinate as a preferred direction. This is expected to influence mainly those structures larger than the turbulence correlation scales. For smaller-scale structures, including the inertial and kinetic ranges, the 
second influence - that of the local large-scale magnetic field - is expected 
to exert a dominant influence. 
Indeed, it is well established
that in the inertial range 
of 
magnetohydrodynamic (MHD) turbulence, the correlation functions
(or equivalently, the spectra) 
of magnetic field and velocity fluctuations exhibit anisotropy relative to the magnetic field direction \citep{MattEA90,MattEA96-var,chen2012ApJ_density_aniso,shaikh2010turbulent,OughtonEA15}.
The symmetries that may be associated with this 
anisotropy may be referred to as rotational symmetries, such as 
axisymmetric ``slab'' or ``2D'' geometries
\citep{BieberEA96}.
The analogous issue of 
anisotropy of density fluctuations 
has received some attention, reviewed below, in the 
theoretical and 
numerical simulation literature \citep{MattEA96-var,ChoLazarian02-prl,Chandran02-radio,ZankEA12-density} 
as well as in remote sensing observations
\citep{ColesHarmon89}.
However, to our knowledge, the issue of correlation or spectral 
anisotropy of the density fluctuation field in the energy-containing and inertial ranges of turbulence
has not been fully examined in solar wind {\it in situ} observations. 
Here we take a step in that direction by examining density correlation statistics,  
their variation relative to the mean magnetic field
and their variation with scale. 
(Mean values and other statistics are computed over samples 
of at least a correlation scale, in accord with 
classical ergodic theory; see e.g., \cite{Panchev}.) 
As in the usual picture of turbulence, scales can be categorized as the energy-containing, inertial, and dissipation regimes. Here, we focus on the former two, which are roughly separated by the correlation scale \citep{frisch1995turbulence}.
Our study emphasizes
observations near 1 au, where long-term datasets provide the possibility of 
high statistical weight 
analyses.

Coronal and solar wind
density fluctuations
can be studied  
based on 
remote sensing techniques, 
such as 
analysis of 
scintillation of signals 
from distant radio
sources 
\citep{ColesHarmon89,ArmstrongEA90,KelloggHorbury05,kontar2023ApJ}.
Rotational symmetry is frequently extracted from these measurements. 
Many of these
studies are 
designed to detect coronal density properties, 
while 
a few have been carried out near 1 au.
The seminal work of \citet{CelnikierEA87}
describes the limitations 
and sensitivities of 
this class of 
scintillation studies.
A typical conclusion is that 
structures in the 
coronal 
density fluctuation field are preferentially elongated
in the
direction of 
the inferred 
mean magnetic field. 
This implies that 
density gradients are stronger in directions perpendicular
to the magnetic field. 
This sense of correlation anisotropy is familiar in solar wind measurements of velocity and magnetic field
\citep{MattEA90,BieberEA96,HamiltonEA08,NaritaEA10-EbAniso, ChenEA11-SWaniso, HorburyEA12,OughtonEA15}.

Interplanetary 
density spectra have 
also been examined based on {\it in situ} observations \citep[see, e.g.][]{BellamyEA05}. But the
directional dependence 
of density fluctuation statistics 
in the 
inertial and energy-containing ranges of scales have sometimes been overlooked, even in relatively complete characterizations of turbulence correlations
\citep{Borovsky12-spectra}.
When density fluctuations have been considered, the 
emphasis has often been on  
high frequency or sub-ion 
scales
\citep{CelnikierEA87,MalaspinaEA10,ChenEA11-SWaniso,kontar2023ApJ}
where kinetic plasma properties are probed.
However, we are not aware
that a comprehensive survey 
has been 
carried out to describe the 
anisotropy of MHD-scale
inertial-range correlation
of the plasma
density. 
This motivates 
our current focus on the anisotropy 
of energy-containing range and 
inertial range density fluctuations near Earth's orbit.

This paper is organized as follows: in Section \ref{sec:theory} we discuss the ``Maltese cross'' representation of correlation anisotropy, which serves as the theoretical basis prompting this research. In Section \ref{sec:data} we describe our data and analysis procedure. Section \ref{sec:results} presents our results on the scale-dependent density correlation anisotropy, and Section \ref{sec:discussion} discusses the implications of the results.

\section{Simplified representations of  anisotropy}
\label{sec:theory}

A point of reference 
that
motivates the present study is the ``Maltese cross'' autocorrelation pattern \citep{MattEA90}
derived from the interplanetary magnetic field (IMF) at 1 au.
The pattern is assumed to be axisymmetric about the mean magnetic field ${\bf B}_0$
and consists of a lobe that admits gradients mainly in the direction
parallel to ${\bf B}_0$ and another part that varies mainly in the directions perpendicular to ${\bf B}_0$. 
In an idealized sense, the former are known as ``slab'' fluctuations, and the latter, 
``2D'' fluctuations.
This so called {\it two-component model}
has become a useful parameterization for anisotropy that incorporates both Alfv\'en wave-like spectral components and a quasi-two-dimensional (2D or Q2D) ingredient that varies, at most, weakly along a mean field \citep{BieberEA94,BieberEA96}. 
The two-component parameterization allows for arbitrary admixtures of energy in models that
vary mainly along or transverse to the mean field, and as such has become a convenient and often-invoked 
model for use in theoretical work on charged 
particle scattering \citep{Zank04, Shalchi, Shalchi10, Zhao17, Zhao18}. 
It is also incorporated with several variations 
into turbulence transport models that 
describe turbulence throughout
the heliosphere as well as models for solar wind acceleration and evolution
in the presence of turbulence \citep[e.g.,][]{AdhikariEA17,UsmanovEA18}. 
The anisotropy present in such models exerts a strong influence on the 
results of such calculations and modeling of turbulence.
It is essential to bear in 
mind that 
such parameterization
of anisotropy are crude representations, and are not intended as dynamical turbulence
models.
However, they demonstrate the physical significance 
and impact of correlation or spectral anisotropy.

Another approach to describing spectral 
anisotropy is based on wave theory, with the premise being that 
linear MHD wave modes may be separated unambiguously based on their polarization properties
\citep{ChoLazarian02-prl, ChoEA02a}.
(The standard decomposition has been controversial for some time and recently a more complete approximate representation that includes structures as well as waves has been suggested by \citet{ZankEA23-linear}.) 
This wave decomposition idea can be directly carried over to turbulence in the {\it weak turbulence} regime.
In that case, the adopted basis and leading-order dynamical solutions are constructed from the linear modes themselves \citep{Chandran05}.
In this view, the Alfv\'en mode is anisotropic, a well-established property in strong MHD turbulence \citep{shebalin1983JPP,OughtonEA94}.
In addition, the Alfv\'en mode 
is polarized transverse to the mean magnetic field, a small-amplitude property adopted in {\it critical balance} theory by 
\citet{GoldreichSridhar95} \citep[see also][]{OughtonMatthaeus20}.
The slow mode is 
assumed to follow a passive dynamics, and to 
admit an anisotropy similar to the
incompressible Alfv\'en mode. 
The decomposition into wave modes 
is completed by extracting 
fluctuations with the polarization of 
linear fast modes. These remain 
isotropic,  as the fast mode dispersion does
not depend on direction.
It is widely regarded
that useful results have been attained based on a
linear-wave decomposition.
However, it should be 
disparaged as a general 
representation of MHD turbulence, 
as it has been shown to be 
essentially incomplete
\citep{ZankEA23-linear};
in particular it lacks 
coherent structures
and nonpropagating structures that are nonetheless found to be dynamically important \citep{GanEA22, Zhao23}.

The linear wave theory 
underlies 
a popular
surrogate for compressional 
effects, the so-called
{\it magnetic compressibility},
that 
has been extensively employed in observational solar wind studies \citep{BrunoCarboneLRSP13,chen2012ApJ_density_aniso}. 
This surrogate
assumes that the relative strength of the 
component of the magnetic variance
parallel to the ambient (mean) 
magnetic field is a measure of compressional 
dynamical activity.
This assumption breaks down 
for large-amplitude turbulence, wherein
parallel fluctuations need not be identified with compressional 
fast magnetosonic modes.
Such fluctuations could
have a more general 
character, such as
an indication of 
spherical polarization
of large-amplitude Alfv\'enic
fluctuations
\citep{BarnesHollweg74,Barnes81}, which 
are usually not associated with 
density variations. 

In the following, we will not make explicit use of representations based on mode decomposition,
but rather will incorporate the underlying ideas into our physical
discussion of the anisotropy of solar wind density fluctuations.

\section{Data and Analysis Procedure}
\label{sec:data}

We acquire 10 years (from February 1998 to March 2008) of ion number density data observed by the SWEPAM instrument on the Advanced Composition Explorer (ACE) spacecraft \citep{McComas98-swepam}, along with corresponding solar wind speed measurements from SWEPAM and magnetic field measurements from the MAG instrument \citep{Smith98-mag} on ACE. This dataset covers most of a solar activity cycle, with maximum sunspot number occurring in 2003 \citep[e.g.,][]{Balmaceda2009JGRA}. The original data at 64-second resolution is upsampled to a 1-minute cadence and separated into datasets that span one day. 
Overlapping
midnight-to-midnight and noon-to-noon intervals are included to increase the total number of datasets and suppress systematic day-timescale periodicities. We remove interplanetary coronal mass ejections (ICMEs) from the data samples, under the assumption that their characteristic average plasma properties 
relegate them to a distinct class of wind intervals \citep{KleinBurlaga82,CaneRichardson03}
that we do not consider here; 
for this, we consulted an available
online table of ICMEs \citep{RichardsonCane24}. About 11\% of intervals with wind speeds below 400 km/s are removed, while approximately 15\% are removed for wind speeds surpassing 400 km/s. The difference may be attributed to the prevalence of ICMEs during the solar maximum, particularly when fast solar winds are more abundant within the ecliptic plane.
In addition, a density interval is discarded if over 70\% of the observations are empty. Within the retained intervals, any missing data - necessarily less than 70\% of the span of the interval – is marked  ``NAN'' to exclude it from our computations. We then compute a linear least squares fit to each interval, and detrend the interval by subtracting the fit from the data, resulting in zero-mean data samples.

We group the data intervals based on their mean solar wind speed $V_\mathrm{SW}$ and the angle between their mean magnetic and velocity fields:
\begin{equation}
\theta = \cos^{-1}{\left(
\frac{\langle \tilde{\bf B} \rangle  \cdot \langle \tilde{\bf V}_\mathrm{SW} \rangle } {\langle|{\bf B}|\rangle \langle|{\bf V}_\mathrm{SW}|\rangle} \right)}
\label{eq:theta}
\end{equation}
where the tilde notation $\tilde{\bf A}$ indicates computing the absolute value of each component of a vector ${\bf A}$, and \(\langle \cdots \rangle\) refers to averaging over an individual 24-hr dataset. Taking the absolute value effectively avoids the cancellation of $\theta$ within a given interval due to magnetic polarity reversals. The $\theta$ channels are $0 - 40$, $40 - 45$, $45 - 50$, $50 - 55$, $55-60$, $60-65$, and $65 - 90$ degrees, and the $V_\mathrm{SW}$ channels are $0-400$, $400-500$, and $500-1000$ km/s for slow, medium, and fast winds, respectively. The channels are chosen to ensure a sufficient number of datasets in each group, as shown in Table~\ref{tab:ACE_cnt} (also see Appendix \ref{sec:app1} for a distribution of the solar wind speeds).

\begin{table}
\centering
  \caption{ACE 24-hour dataset count in each wind speed and angular channel. The angle $\theta$ is defined in Eq.~\ref{eq:theta}}
\begin{tabular}{@{}c|ccc}
  \hline
   & $\leq 400$ km/s & 400-500 km/s & $\geq 500$ km/s \\
  \hline
  $0^\circ$-$40^\circ$ & 62 & 81 & 34\\
  $40^\circ$-$45^\circ$ & 96 & 97 & 77\\
  $45^\circ$-$50^\circ$ & 206 & 283 & 284\\
  $50^\circ$-$55^\circ$ & 341 & 487 & 563\\
  $55^\circ$-$60^\circ$ & 449 & 621 & 492\\
  $60^\circ$-$65^\circ$ & 373 & 400 & 227\\
  $65^\circ$-$90^\circ$ & 392 & 265 & 81\\
  \hline
  Total & 1919 & 2234 & 1758\\
  \hline
\end{tabular}
\label{tab:ACE_cnt}
\end{table} 

To proceed with our analysis, 
we compute the density autocorrelation function for each dataset using the Blackman-Tukey method \citep{BlackmanTukey} (as described in detail in \citet{RoyEA21}). 
The ensemble definition of the autocorrelation is
\begin{equation}
R(\tau) = \langle \rho(t) \rho(t+\tau)\rangle - 
\langle \rho(t) \rangle \langle \rho(t+\tau)\rangle
\label{BT}
\end{equation}
where \(\rho\) is the ion density, $\tau$ is the time lag. Invoking the ergodic theorem, the brackets $\langle \cdots \rangle$
correspond formally to averaging over an
infinite sample size. For finite data consisting of $N$ equally spaced samples, we denote the averaging operation as $\langle \cdots \rangle'$.
Specifically, for a dataset $\{\rho_i\} = \rho_0, \cdots, \rho_{N-1}$ with sampling time $\Delta t=60$ seconds, $\tau$ takes integer multiples of $\Delta t$, and the averaging can be written explicitly as
\begin{equation}
    \langle \rho(t) \rangle' = \langle \rho_j \rangle_{j=0, \cdots, N-\tau/\Delta t-1},
\end{equation}
\begin{equation}
    \langle \rho(t+\tau) \rangle' = \langle \rho_j \rangle_{j=\tau/\Delta t, \cdots, N-1},
\end{equation}
\begin{equation}
    \langle \rho(t) \rho(t+\tau) \rangle' = \langle \rho_j \rho_{j+\tau/\Delta t} \rangle_{j=0, \cdots, N-\tau/\Delta t-1}.
\end{equation}
For the remainder of the paper, for clarity, we drop the
prime in the bracket notation. 

For stationary data, $R(\tau)$ does not depend on the variable $t$, i.e. the origin of time. And therefore by definition, the correlation function is an even function of lag $\tau$. Under appropriate conditions, this is equivalent to
the Reynolds averaging expression
for the correlation function, $R(\tau) = \langle \delta \rho(t) \delta \rho(t+\tau)\rangle $, where 
$\delta \rho(t)  \equiv \rho (t) - \langle \rho(t)\rangle$ \citep[see][]{Germano92}. 

When we consider the 
normalized correlation function $\hat{R}$, the $R(\tau$) resulting from Eq.~\ref{BT} is 
normalized by the data variance $R(0)$:
\begin{equation}
    \hat{R}(\tau) = \frac{R(\tau)}{R(0)}.
\end{equation}

To eliminate undersampled fluctuations at large lags, we pass each autocorrelation function through a 10\% cosine taper window \citep{matthaeus1982measurement}, where the last 10\% of $\hat{R}(\tau)$ values are multiplied by the factor
\begin{equation}
    \frac{1}{2} \left( 1+\cos{ \left[ \frac{\pi}{0.1 \tau_\mathrm{max}} \left( \tau-0.9\tau_\mathrm{max}\right) \right]} \right)
\label{cosinetap}
\end{equation}
with $\tau_\mathrm{max}$ representing the maximum lag over which the autocorrelation is calculated. In this analysis the maximum lag is 4.8 hours; this corresponds to $1/5$ of the data interval, and 
several times the anticipated correlation times. 

We further transform the temporal lags $\tau$ into spatial lags $\lambda$ by applying the Taylor frozen-in hypothesis \citep{Taylor38}, $\lambda = - V_\mathrm{SW} \tau$, where $V_\mathrm{SW}$ is the solar wind speed in the upstream direction averaged over the data interval. With this procedure, we arrive at a normalized, spatial lag-dependent correlation function
$\hat R(\lambda)$.

%In a similar way we can compute the second order structure function,
%\begin{equation}
%    D^{(2)}(\lambda) = \langle (\delta \rho(x + \lambda) - \delta \rho(x))^2 \rangle.
%    \label{eq:S2}
%\end{equation}
%The structure
%function \victoria{of a stationary random process has a one-to-one correspondence with the correlation function, }
%$D^{(2)}(\lambda) = 2R(0) - 2 R(\lambda)$,
%has advantages 
%in rates of convergence at small lags.
%This is mainly 
%due to 
%information about the total variance, $\delta \rho^2 = R(0)$ the mean square density fluctuation, being transformed to asymptotically large lags, rather than appearing at small lags as in the correlation function itself.
%Note that due to the finite length of individual data samples, there is necessarily a small differences in the data employed to compute the ensemble statistics of $\langle \delta \rho^2 (x+\lambda) \rangle$ and the statistics of $\langle \delta \rho^2 (x) \rangle$.

%An equivalent spectrum \citep{ChhiberEA18} may also be conveniently computed from the second order structure function. This quantity is so named because it behaves similarly to the wavenumber spectrum in a range of scales wherein the signal has a power-law spectrum. We may define it as
%\begin{equation}
%    S_{eq}(k=1/\lambda) =
%     \lambda D^{(2)}({\lambda}),
%    \label{eq:Seq}
%\end{equation}
%where $k$ is an estimated equivalent wavenumber.

\section{Results}
\label{sec:results}

\begin{figure}
\centering
    \includegraphics[angle=0,width=0.9\columnwidth]{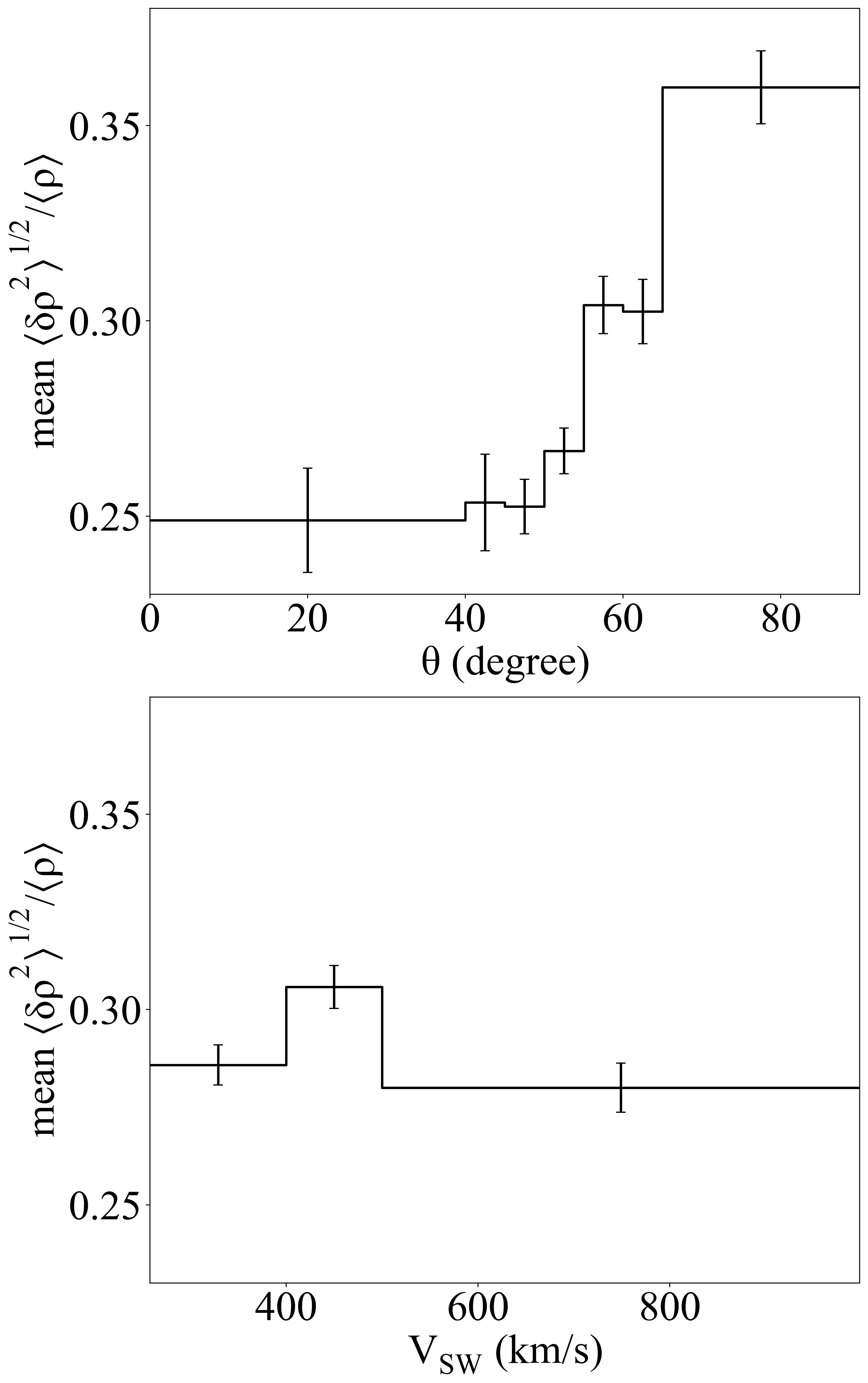}
    \caption{Density coefficient of variation averaged over each $\theta$ channel (top panel) and each $V_\mathrm{SW}$ channel (bottom panel). Error bars represent standard errors. Here $\delta \rho(t)  \equiv \rho (t) - \langle \rho(t)\rangle$, where \(\langle \cdots \rangle\) refers to an average over an individual 24-hr dataset, and ``mean'' refers to averaging the coefficient of variations over all 24-hour datasets that lie within a \(\theta\) or \(V_\mathrm{SW}\) channel.}
\label{fig:var_dist}
\end{figure}

We first examine statistics of the density samples. 
The standard deviation over mean, otherwise known as the coefficient of variation, C.V., is a measure of the relative fluctuation amplitude of the density samples. The C.V. range up to $1.57$ across the 24-hour datasets and increase with $\theta$. 
Fig.~\ref{fig:var_dist} illustrates 
the average sample C.V. with respect to $\theta$ and $V_\mathrm{SW}$, with the error bars representing standard errors. These plots indicate that relative density fluctuations are stronger when the spacecraft samples flow perpendicular to, rather than parallel to, the mean magnetic field, and that slow wind exhibits slightly larger density fluctuations compared to fast wind. The top panel of Fig.~\ref{fig:var_dist} is consistent with the recent work of \cite{Du2023ApJ}, on the anisotropy of density fluctuations obtained from 
simulations of compressible MHD turbulence. We present further statistics of our density samples in Appendix~\ref{sec:app2}, which includes the standard deviation $\langle \delta \rho^2 \rangle^{1/2}$ of density for each angular and wind speed channel. It is well-known that slow wind is denser than fast wind \citep{McComasEA00,UsmanovEA18}; it is shown in Appendix~\ref{sec:app2} that the density fluctuation magnitudes in slow wind are 
also larger. We also include a joint
distribution of the density standard deviation and mean.

\begin{figure}
\centering
    \includegraphics[angle=0,width=\columnwidth]{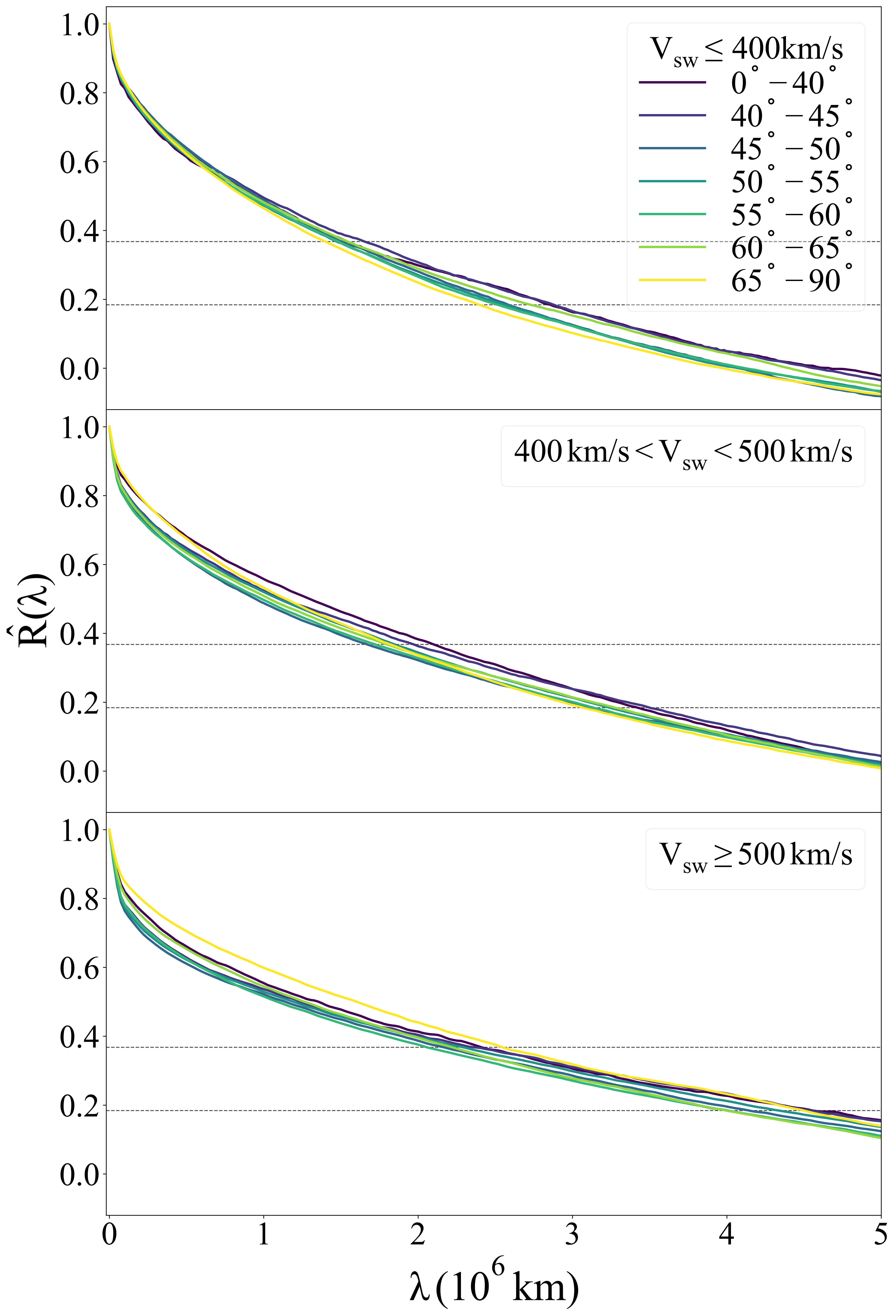}
    \caption{Normalized density autocorrelation for slow (top panel), medium (middle panel), and fast (bottom panel) winds as functions of spatial lag. The correlations have been averaged within each angular channel, denoted by distinctly colored lines. Dashed horizontal lines indicate where $\hat{R} = 1/e$ and $1/2e$.}
\label{fig:correlationAvg}
\end{figure}

We proceed to our investigation of the correlation anisotropy. The normalized autocorrelation functions are averaged within each group of speed and angular channel, resulting in 21 instances of autocorrelations henceforth represented by $\hat R(\lambda = -V_\mathrm{SW} \tau)$. These are shown in the 
panels of Fig.~\ref{fig:correlationAvg}.
Separate panels correspond to slow, medium, and fast winds, and within each, angular variations are demonstrated.

\begin{figure}
\centering
    \includegraphics[angle=0,width=\columnwidth]{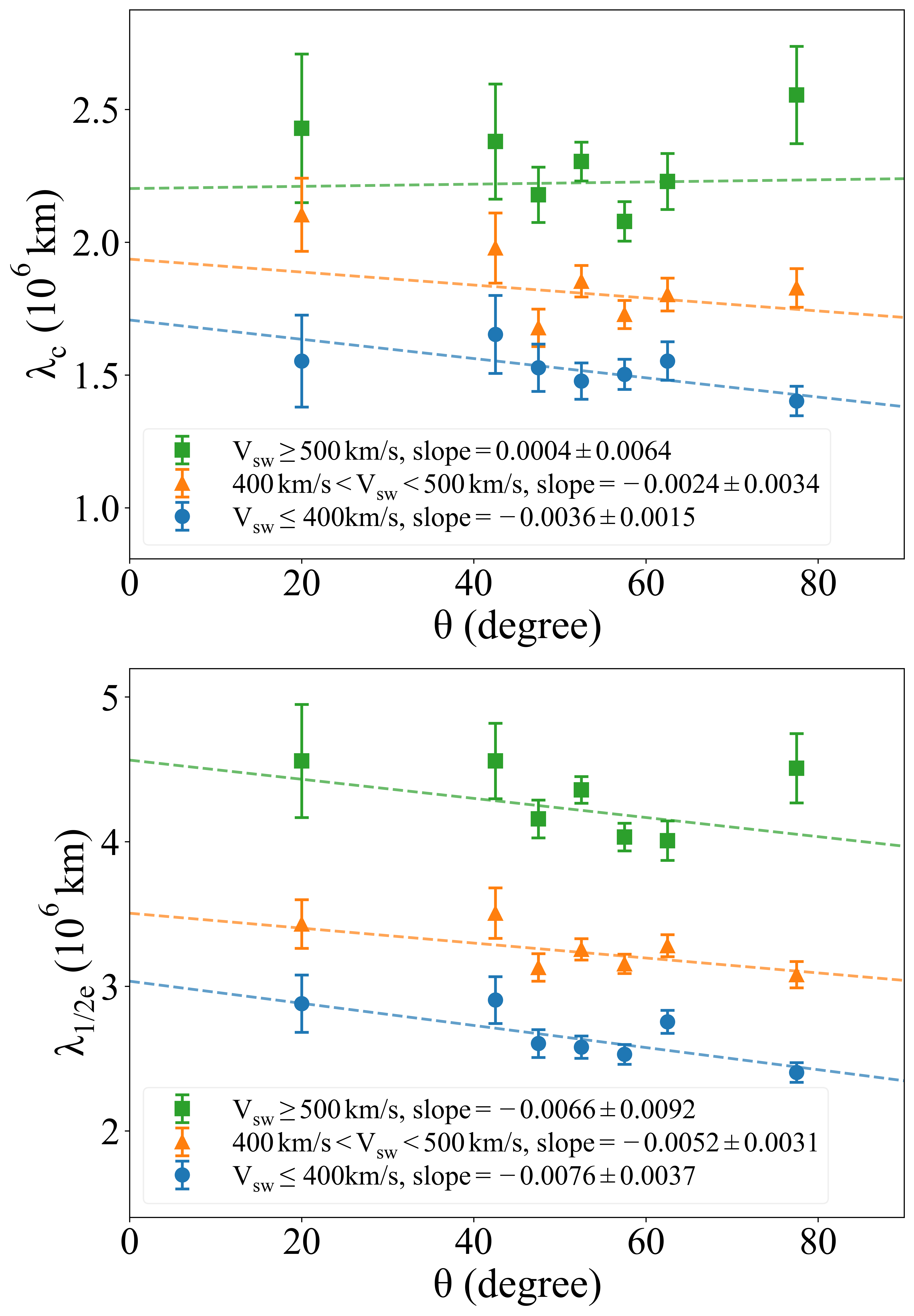}
    \caption{The correlation length $\lambda_c$ (top panel) and the length $\lambda_{1/2e}$ where the correlation decreases by a factor of $1/2e$ (bottom panel) as functions of angular channel for slow, medium, and fast winds. The data and uncertainties are consistent with those listed in Table~\ref{tab:Lc}. Dashed lines show linear best fits with corresponding slopes listed in the legends.}
\label{fig:theta_Lc}
\end{figure}

The correlation length $\lambda_c$, defined as the spatial lag where the correlation decreases by a factor of $1/e$, is listed in Table~\ref{tab:Lc} for each channel. The uncertainties are standard errors calculated from the ensemble of intervals within each channel. We further plot $\lambda_c$ over $\theta$ for all three wind channels in the top panel of Fig.~\ref{fig:theta_Lc}, and fit each set of data with a linear curve using least squares analysis. We find a general subtle trend of correlation length decreasing as $\theta$ increases in both slow and medium wind - the slope of slow wind is negative with a $2\sigma$ confidence, as shown in the legend of Fig.~\ref{fig:theta_Lc}. However, the slope of fast wind is close to zero. This indicates that the longest average correlation lengths in slow and medium winds occur when the mean magnetic field direction is quasi-aligned with the plasma flow direction, while in fast wind the correlation lengths are approximately isotropic.

\begin{table}
\centering
  \caption{Correlation length from the $1/e$ method and the length where correlation decreases by $1/2e$ (in parentheses) in units of $10^6$ km in each wind speed and angular channel. The uncertainties represent standard errors.}
\begin{tabular}{@{}c|ccc@{}}
  \hline
   & $\leq 400$ km/s & 400-500 km/s & $\geq 500$ km/s \\
  \hline
  {\multirow{2}{*}{$0^\circ$-$40^\circ$}} & $1.55\pm .17$ & $2.10\pm .14$ & $2.43 \pm .28$\\ 
  & ($2.88 \pm .20$) & ($3.43 \pm .17$) & ($4.56 \pm .39$)\\
  {\multirow{2}{*}{$40^\circ$-$45^\circ$}} & $1.65\pm .15$ & $1.98\pm .13$ & $2.38 \pm .22$\\
  & ($2.90 \pm .16$) & ($3.51 \pm .17$) & ($4.56 \pm .26$)\\
  {\multirow{2}{*}{$45^\circ$-$50^\circ$}} & $1.53\pm .09$ & $1.68\pm .07$ & $2.18\pm .10$ \\
  & ($2.60 \pm .10$) & ($3.13 \pm .10$) & ($4.16 \pm .13$)\\
  {\multirow{2}{*}{$50^\circ$-$55^\circ$}} & $1.48\pm .07$ & $1.85 \pm .06$ & $2.30 \pm .07$ \\
  & ($2.58 \pm .08$) & ($3.26 \pm .07$) & ($4.36 \pm .09$)\\
  {\multirow{2}{*}{$55^\circ$-$60^\circ$}} & $1.50\pm .06$ & $1.73 \pm .05$ & $2.08 \pm .07$\\
  & ($2.53 \pm .07$) & ($3.16 \pm .07$) & ($4.03 \pm .10$)\\
  {\multirow{2}{*}{$60^\circ$-$65^\circ$}} & $1.55\pm .07$ & $1.80 \pm .06$ & $2.23 \pm .11$\\
  & ($2.75 \pm .08$) & ($3.28 \pm .08$) & ($4.01 \pm .14$)\\
  {\multirow{2}{*}{$65^\circ$-$90^\circ$}} & $1.40\pm .06$ & $1.83 \pm .07$ & $2.55 \pm .18$\\
  & ($2.40 \pm .07$) & ($3.08 \pm .09$) & ($4.51 \pm .24$)\\
  \hline
  {\multirow{2}{*}{Average}} & $1.52\pm .27$ & $1.85\pm .24$ & $2.31 \pm .44$ \\
  & ($2.67 \pm .31$) & ($3.26 \pm .30$) & ($4.31 \pm .58$)\\
  \hline
\end{tabular}
\label{tab:Lc}
\end{table} 

We also list, in parentheses in Table~\ref{tab:Lc}, the spatial lag where the correlation decreases by a factor of $1/2e$, denoted as $\lambda_{1/2e}$. The corresponding plot is shown in the bottom panel of Fig.~\ref{fig:theta_Lc}. We find that as compared to $\lambda_c$, $\lambda_{1/2e}$ decreases more noticeably as $\theta$ increases, suggesting that the observed correlation persists to larger distances in the parallel directions. Equivalently, the gradients at the outer scale of turbulence are moderately stronger in the perpendicular directions. Additionally, we observe that the correlation lengths are systematically longer in fast wind compared to slow wind. This latter difference appears to be clearer in density correlations compared with magnetic correlations \citep{WeygandEA11}.

\begin{figure}
\centering
    \includegraphics[angle=0,width=0.85\columnwidth]{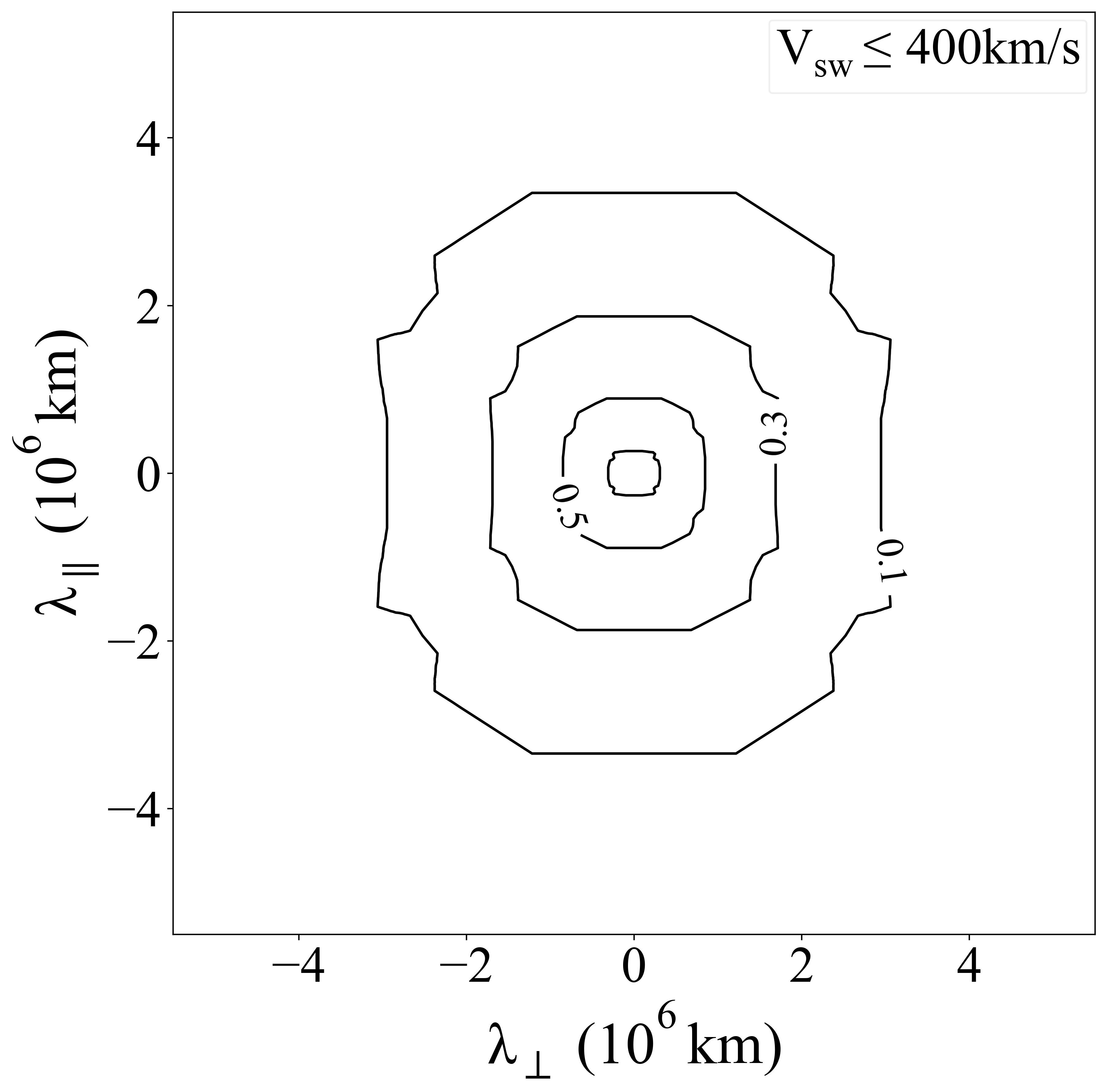}
    \includegraphics[angle=0,width=0.85\columnwidth]{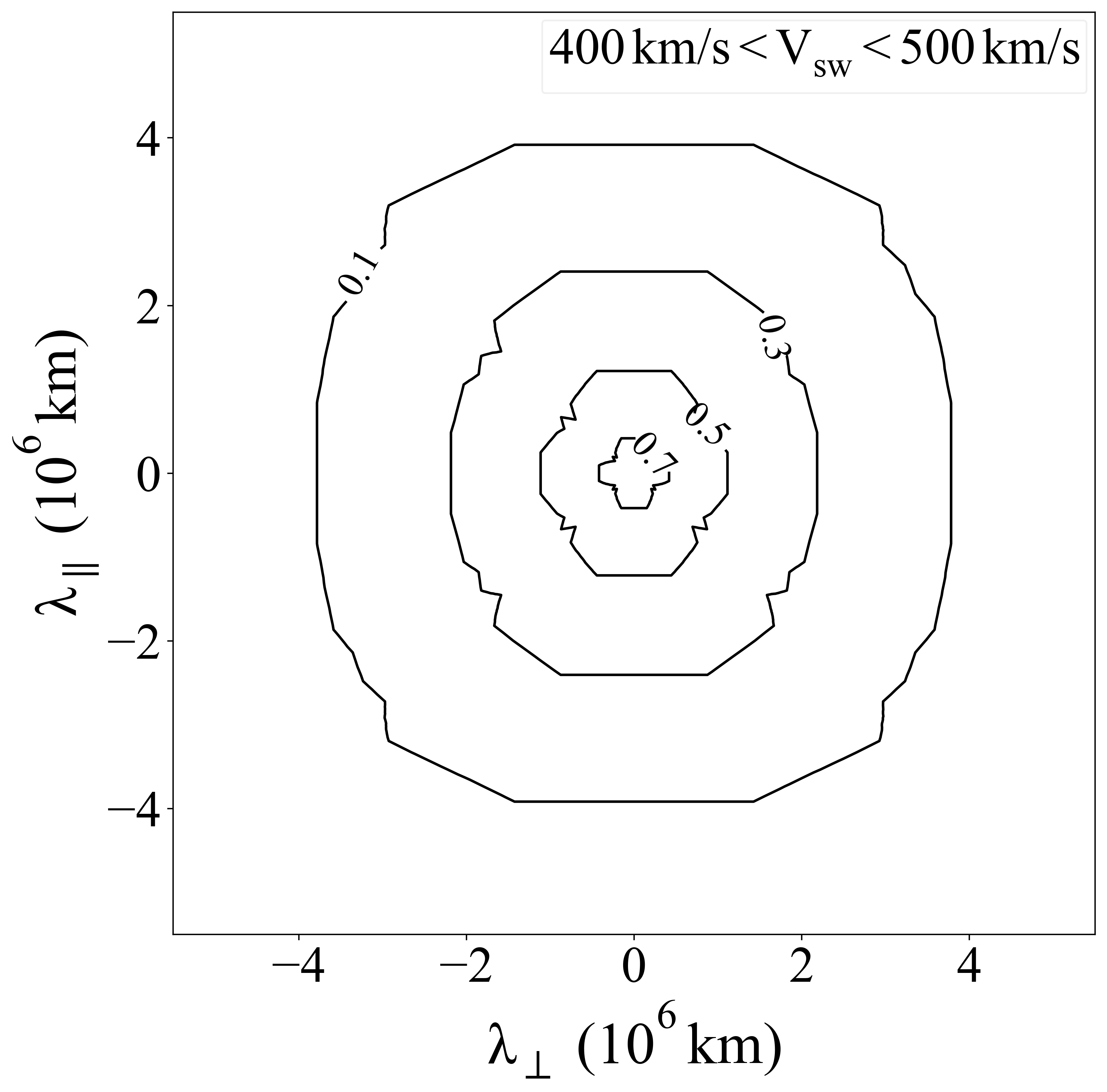}
    \includegraphics[angle=0,width=0.85\columnwidth]{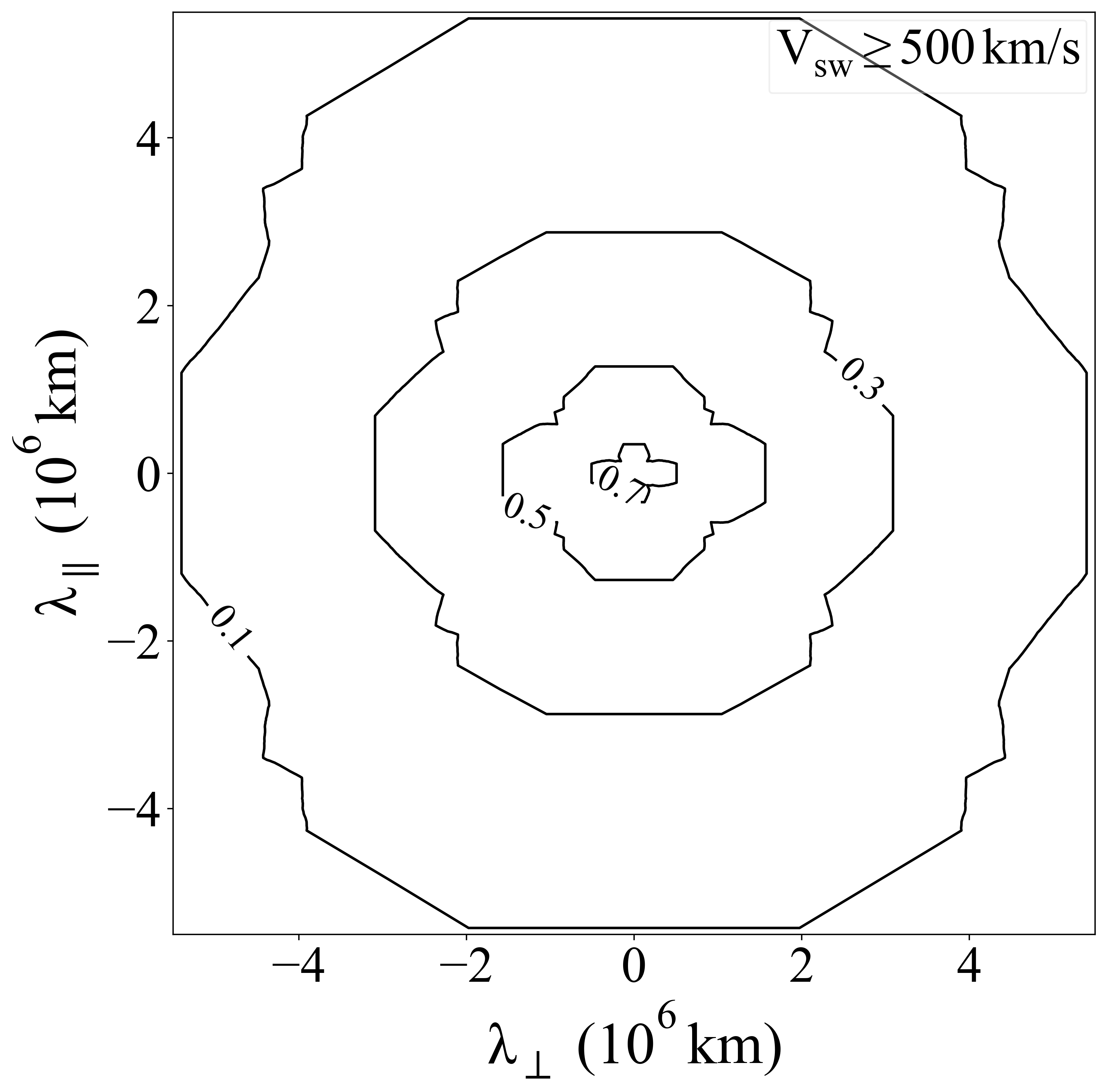}
    \caption{The correlation contours for slow (top panel), medium (medium panel), and fast (bottom panel) winds, calculated from Fig.~\ref{fig:correlationAvg} through the transformation $(\lambda_\perp = \lambda \sin{\theta}, \lambda_\parallel = \lambda \cos{\theta})$.} 
\label{fig:Mcrosss}
\end{figure}

To better visualize the density correlation anisotropy for slow, medium, and fast winds, 
in Fig.~\ref{fig:Mcrosss}
we plot the contour levels of the averaged autocorrelations $\hat{R}(\lambda)$ in perpendicular and parallel lag spaces through the transformation $(\lambda_\perp = \lambda \sin{\theta}, \lambda_\parallel=\lambda\cos{\theta})$, following \citet{DassoEA05}. The contours are computed in the first quadrant, then mirrored 
about the $\theta=0^\circ$ and $90^\circ$ axes under symmetry assumptions. This is a statistical demonstration of the``Maltese cross'' geometry in density fluctuation fields.

It is evident that the outermost contours of constant correlation in Fig.~\ref{fig:Mcrosss}, those with the smallest correlation values, are slightly elongated in the parallel direction for slow and medium wind speed. These contours of 
very low correlation values (the largest ``circles'') 
may be characterized as weakly 
``2D-like'' in the sense described in 
\citet{DassoEA05}, whereas the fast wind contour is isotropic.
However, the higher correlation contours (the smaller circles) tend to exhibit the opposite type of anisotropy that may be described as ``slab-like''. 
Specifically, for fast wind, the contours with correlation greater than $0.1$ is somewhat ``slab-like" with elongation in the perpendicular direction, and this tendency becomes more pronounced
at higher correlation values, as depicted in the bottom panel of Fig.~\ref{fig:Mcrosss}.  
Meanwhile, the $\hat{R} = 0.5$ contours for slow and intermediate winds remain slightly 2D-like, even though they also become slab-like at the 
highest correlation values. 
This general tendency of the correlation contours in its dependence on wind speed is qualitatively consistent with the magnetic field results from \citet{DassoEA05}.  
We discuss this comparison in greater detail later. 

The general 
increase in correlation lengths
with increasing wind speed is also readily apparent in Fig.~\ref{fig:Mcrosss}, where isocontours ``expand'' from the top to the bottom panel.

\begin{figure}
\centering
    \includegraphics[angle=0,width=\columnwidth]{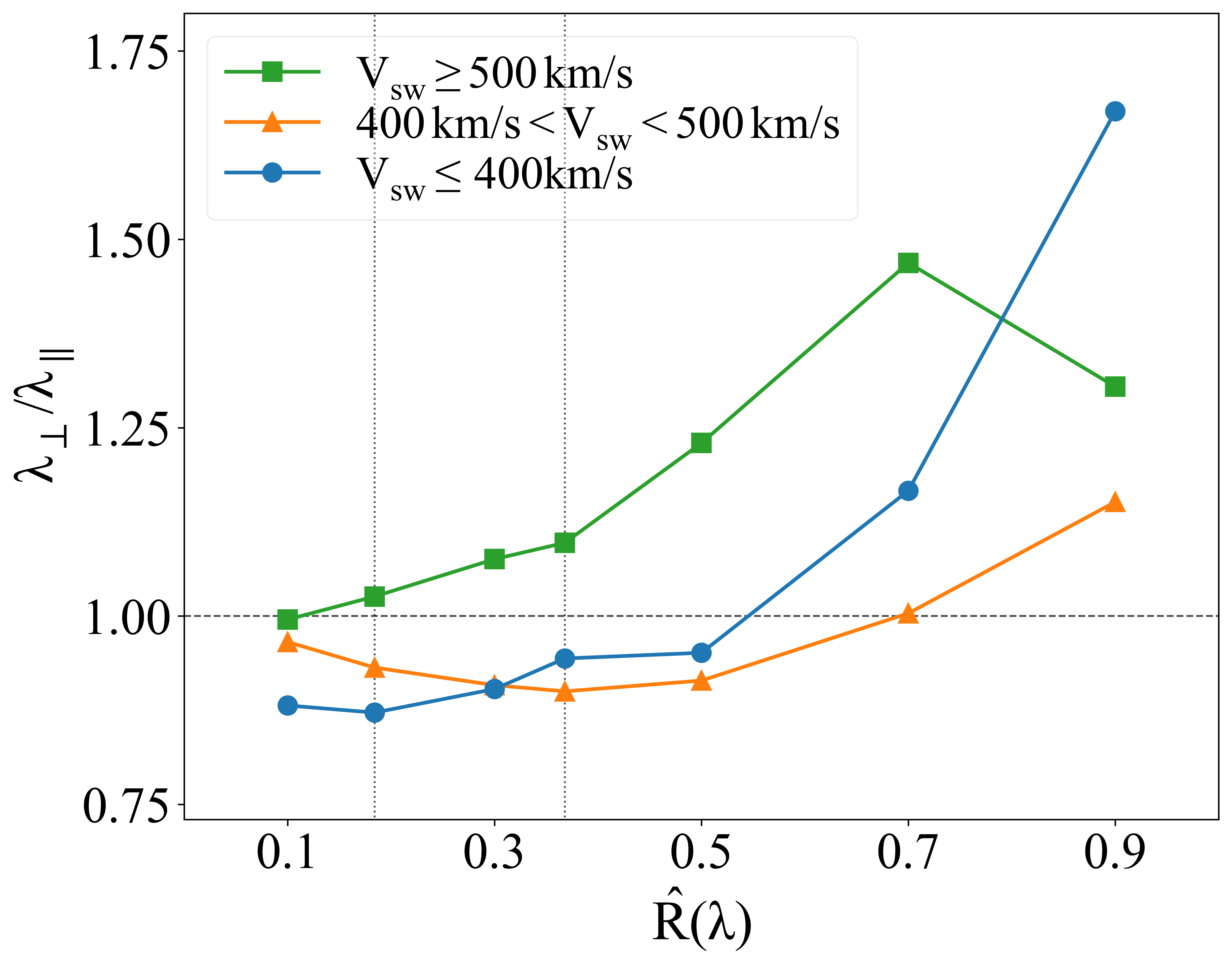}
    \caption{Estimates of $\lambda_\perp/\lambda_\parallel$ as functions of $\hat{R}(\lambda)$ for slow, medium, and fast winds. Dotted vertical lines indicate the $1/e$ and $1/2e$ correlation levels. Dashed horizontal line indicates where $\lambda_\perp/\lambda_\parallel = 1$.}
\label{fig:lambdas}
\end{figure}

Table~\ref{tab:anisotropy}
quantifies the density anisotropies as we have 
described above. 
Here, we list the values for $\lambda_\perp$, $\lambda_\parallel$, and $\lambda_\perp/\lambda_\parallel$ in all three wind speed channels corresponding the following normalized correlations: $\hat{R} = 0.9$, 0.7, 0.5, $1/e$, 0.3, $1/2e$, and 0.1. 
Scales larger than the correlation scale are considered the energy-containing scale, while smaller scales belong to the inertial range. The lags are calculated from Fig.~\ref{fig:correlationAvg} using $\lambda_\perp = \lambda(\hat{R}_{65^\circ-90^\circ}) \sin{77.5^\circ}$ and $\lambda_\parallel = \lambda(\hat{R}_{0^\circ-40^\circ}) \cos{20^\circ}$, or can be directly observed in Fig.~\ref{fig:Mcrosss}. Note that due to the coarse binning at low $\theta$, $\lambda_\parallel$ may be underestimated and $\lambda_\perp/\lambda_\parallel$ may be overestimated. Fig.~\ref{fig:lambdas} shows how $\lambda_\perp/\lambda_\parallel$ varies with $\hat{R}$. 
We again observe that the correlations are slightly elongated along the parallel direction at small $\hat{R}$ (large spatial lags), while for small spatial lags (large $\hat{R}$), corresponding to the inertial range, the correlations are elongated in the perpendicular direction. The systematic underestimation of parallel correlation lengths suggests that the dynamics may be more 2D-like than predicted in Fig.~\ref{fig:Mcrosss} and~\ref{fig:lambdas}.

\begin{table}
\centering
  \caption{Estimates of $\lambda_\perp$ and $\lambda_\parallel$ in units of $10^6$ km as well as their ratio on the correlation contours $\hat{R}(\lambda) = 0.9$, 0.7, 0.5, $1/e$, 0.3, $1/2e$, and 0.1 for slow, medium, and fast winds.}
\begin{tabular}{c|c|ccc}
   \hline
   & $\hat{R}$ & $\lambda_\perp$ & $\lambda_\parallel$ & $\lambda_\perp/\lambda_\parallel$\\
  \hline
  {\multirow{7}{*}{$\leq 400$ km/s}} 
  & 0.9 & 0.04 & 0.02 & 1.67\\
  & 0.7 & 0.31 & 0.26 & 1.17\\
  & 0.5 & 0.85 & 0.89 & 0.95\\
  & $1/e$ & 1.37 & 1.45 & 0.94\\
  & 0.3 & 1.69 & 1.87 & 0.90\\
  & $1/2e$ & 2.34 & 2.69 & 0.87\\
  & 0.1 & 2.95 & 3.34 & 0.88\\
  \hline
  {\multirow{7}{*}{400-500 km/s}} 
  & 0.9 & 0.05 & 0.04 & 1.15\\
  & 0.7 & 0.42 & 0.42 & 1.00\\
  & 0.5 & 1.11 & 1.22 & 0.91\\
  & $1/e$ & 1.78 & 1.97 & 0.90\\
  & 0.3 & 2.18 & 2.41 & 0.91\\
  & $1/2e$ & 3.00 & 3.22 & 0.93\\
  & 0.1 & 3.78 & 3.92 & 0.97\\
  \hline
  {\multirow{7}{*}{$\geq 500$ km/s}} 
  & 0.9 & 0.05 & 0.04 & 1.30\\
  & 0.7 & 0.51 & 0.35 & 1.47\\
  & 0.5 & 1.57 & 1.27 & 1.23\\
  & $1/e$ & 2.49 & 2.27 & 1.10\\
  & 0.3 & 3.09 & 2.87 & 1.08\\
  & $1/2e$ & 4.39 & 4.28 & 1.03\\
  & 0.1 & 5.40 & 5.42 & 1.00\\
  \hline
\end{tabular}
\label{tab:anisotropy}
\end{table}

\section{Discussion}
\label{sec:discussion}

There is a general tendency for 
MHD, and plasma turbulence
in the MHD range of scales as well, 
to exhibit correlation anisotropy, 
and equivalently, spectral anisotropy relative to the direction of a regional mean magnetic field of sufficient strength
\citep{shebalin1983JPP,OughtonEA94}.
Other effects may also introduce anisotropies by imposing 
preferred directions 
that influence regional and local dynamics. 
For example, in a structured and expanding medium such as the solar wind, large-scale plasma flows may introduce preferred directions that influence anisotropy. 
A notable effect is that of 
expansion, which, in the simplest case, selects the 
radial direction as preferred. 
Other relevant 
effects include regions of shear and compression
occurring between 
high- and low-speed streams
and near shocks, as well as interactions 
involving structures such as coronal mass ejections.
One may expect that 
these 
various influences on anisotropy may operate at 
different length scales,
and may have varying levels of 
influence on different 
physical quantities. 
For mostly practical reasons, solar wind studies on turbulence anisotropy have often concentrated on the 
magnetic field. The present
report extends this discussion to include the fluctuations in density. 

More specifically, 
most
previous examinations of anisotropy have considered the relatively local 
effects of the magnetic field 
direction on the rotational symmetry 
of magnetic fluctuations
\citep{BieberEA94,BieberEA96}.
Based on results from 
laboratory experiments
\citep{RobinsonRusbridge71}
and numerical simulations
\citep{shebalin1983JPP,OughtonEA94}, the expectation is that MHD-scale 
turbulence will display 
a quasi-2D anisotropy relative to the field direction. This expectation is 
mainly motivated by a property of  
incompressive dynamics, namely that the cascade to higher perpendicular wavenumbers proceeds unabated while the parallel cascade is inhibited by wave propagation \citep{shebalin1983JPP}. 
This anisotropy is reasonably well-confirmed
in most analyses of solar wind rotational symmetry
\citep{BieberEA94,BieberEA96}.
An exception is 
the study of
\citet{SaurBieber99}, which finds some support for a preferred 
role of the radial direction at relatively lower frequencies.
The later finding suggests the influence of expansion, 
an effect clearly seen in WKB treatments \citep[e.g.][]{VolkAlpers73}.
 
Considering a 
broader context, 
the dynamics 
in the solar wind at 1 au may consist of 
an admixture of 
incompressive and
compressive parts. In this 
more realistic
portrayal, more complex influences on rotational symmetry may be anticipated. 
In particular, 
the incompressive tendency towards a quasi-2D configuration 
merges with the relatively isotropic 
spectral (and correlation) statistics
attributed to the compressible dynamics.
Indeed, compressible MHD
simulations \citep{MattEA96-var,Du2023ApJ}
with Mach number, plasma $\beta$, and 
$\delta B/B$ similar to those of the solar wind indicate that 
density spectra are anisotropic, 
but less so than the 
anisotropy seen in incompressible
simulations \citep{OughtonEA94}.
This appears to be consistent with the current findings in slow and medium winds, 
wherein the correlation lengths parallel and perpendicular to the mean magnetic field differ from one another by only a small relative fraction. 
The present finding 
also suggests a more complex 
scenario in which the 
sense of anisotropy varies across scales. 

It has been previously suggested \citep{DassoEA05,WeygandEA11} that 
for the magnetic field in the solar wind, 
the parallel correlation scale 
is larger than the perpendicular 
correlation scale by a factor of around 2
for the slow solar wind.  
This is a modest 2D-like
anisotropy compared to
what is expected at smaller scales in the inertial range. However, such a ratio 
is considerably larger  
than that of the density
correlation scales reported here. 

Here, for density fluctuations, the 
sense of 
anisotropy at the correlation scale, whether measured at the $1/e$ or the $1/2e$ level, 
is mainly of the ``2D'' type. (The fast wind is essentially isotropic at the $1/e$ scale but becomes slightly 2D at larger scales.) We suggest that the weaker outer-scale anisotropy of the 
density is due to the 
effects of 
the admixture of
more anisotropic incompressive 
turbulence with less anisotropic compressible turbulence, consistent with previous numerical results \citep[see, e.g., Fig.~5 of][]{MattEA96-var}.

It is interesting,
perhaps a bit surprising and of potential significance,
that at smaller scales (higher correlation values), the sense of density anisotropy reverses and favors slab-like symmetries.
This trend occurs in all wind speed channels 
and is especially dramatic for fast
wind. 
This is reminiscent of the sense of magnetic 
anisotropy at correlation scales in fast wind \citep{DassoEA05}. Furthermore, by inspection of the magnetic correlation contours in \citet{DassoEA05}, it appears that slab-like symmetry 
occurs across a wide range of scales in fast wind. This however was not quantified.
But for slow wind, the 2D-like 
sense of magnetic anisotropy mentioned above as present at the correlation scale remains (as seen by inspection) 2D-like over a reasonably wide range of scales. 
Furthermore, in the above-quoted compressible MHD 
simulation results
\citep{MattEA96-var}, the inertial range contours of density spectra appear to be of the 2D type, although not dramatically so. 
Finally, we note that 
our finding of slab-like density in the fast wind inertial range
also 
seems to contradict Fig.~5 and~6 of \citet{chen2012ApJ_density_aniso}, 
who use Ulysses data and 
adopt \(|{\bf B}|\) as a proxy for compressive fluctuations. A major difference, however, is that the analysis of
\citet{chen2012ApJ_density_aniso}
is carried out in a 
coordinate system based on 
a local definition of the mean field. 
Such a procedure 
systematically increases
the ratio of perpendicular to 
parallel structure functions, thus favoring 2D-like interpretations
\citep{MatthaeusEA12}. 
Our computation of mean fields 
integrated over longer times is chosen to 
avoid this bias. 

We cannot rule out the appearance
of field aligned (2D-like) anisotropies
at much smaller scales, possibly for all wind speeds. Indeed, these are favored 
by coronal observations such as \citet{ArmstrongEA90}. The study found field-aligned elongated structures having 
anisotropy ratios that increase with increasing heliocentric distance from 2 to about 10 solar radii. 
However, these observations were at much smaller scales, and 
much closer to the sun, relative to the present 
large-scale observations at 1 au.
Nevertheless, these authors did suggest that solar wind density anisotropy varies with scale. On the other hand, \citet{Zank24arXiv}, studied density fluctuations in sub-Alfv\'enic wind using mode-decomposition analysis, and found anisotropy dominated by a combination of the slab-like entropy mode and the 2D-like backward propagating slow magnetosonic mode. 
The anisotropy of their entropy mode is consistent with our results, although their results extend over the scales $k = 2 \times 10^{-6}$ to $10^{-2}$ km$^{-1}$, which slightly overlaps with the smallest scales we investigate.

The present results for the anisotropy 
of density stand in substantial contrast to expectations based on magnetic and velocity field spectra in incompressible
simulations and in solar wind observations. The basis of this expectation is that the incompressible cascade, which presumably is a major factor in the solar wind dynamics, is well known to favor 2D-like anisotropies \citep{OughtonEA15}.
The reasons for this departure remain unclear at present, but most likely pertain to the way compressible fluctuations are generated in the solar wind. Further research will
be required to arrive 
at a clearer understanding. 

Future observations from the PUNCH mission \citep{deforest2022PUNCH} will provide us with solar wind density data in regions of the inner heliosphere yet unexplored and with an unprecedented field of view. As observed in \citet{deforest2016fading}, the solar wind shows a transition from ``striated'' to ``flocculated'' features, suggesting an evolution toward isotropization \citep{cuesta2022isotropization}. Using the white-light images obtained from PUNCH, it will be possible to recover unprecedented mapping of solar wind density. Such measurements will be used to perform analyses similar to those presented in this paper, that will provide invaluable knowledge about the radial evolution of solar wind anisotropy.

% \begin{acknowledgments}
The velocity and density data were downloaded from \url{https://spdf.gsfc.nasa.gov/pub/data/ace/swepam/level2_hdf/ions_64sec}. This research is partially supported by the NASA LWS grant 80NSSC20K0377 (subcontract 655-001), and by the NASA IMAP project at UD under subcontract SUB0000317 from Princeton University. R.C. and M.E.C. acknowledge LANL's hospitality during Summer 2022, when part of this work was performed.
% \end{acknowledgments}
\clearpage

\appendix
\section{Distribution of solar wind speed at 1 au}
\label{sec:app1}
\begin{figure}[!htbp]
\centering
    \includegraphics[angle=0,width=0.6\columnwidth]{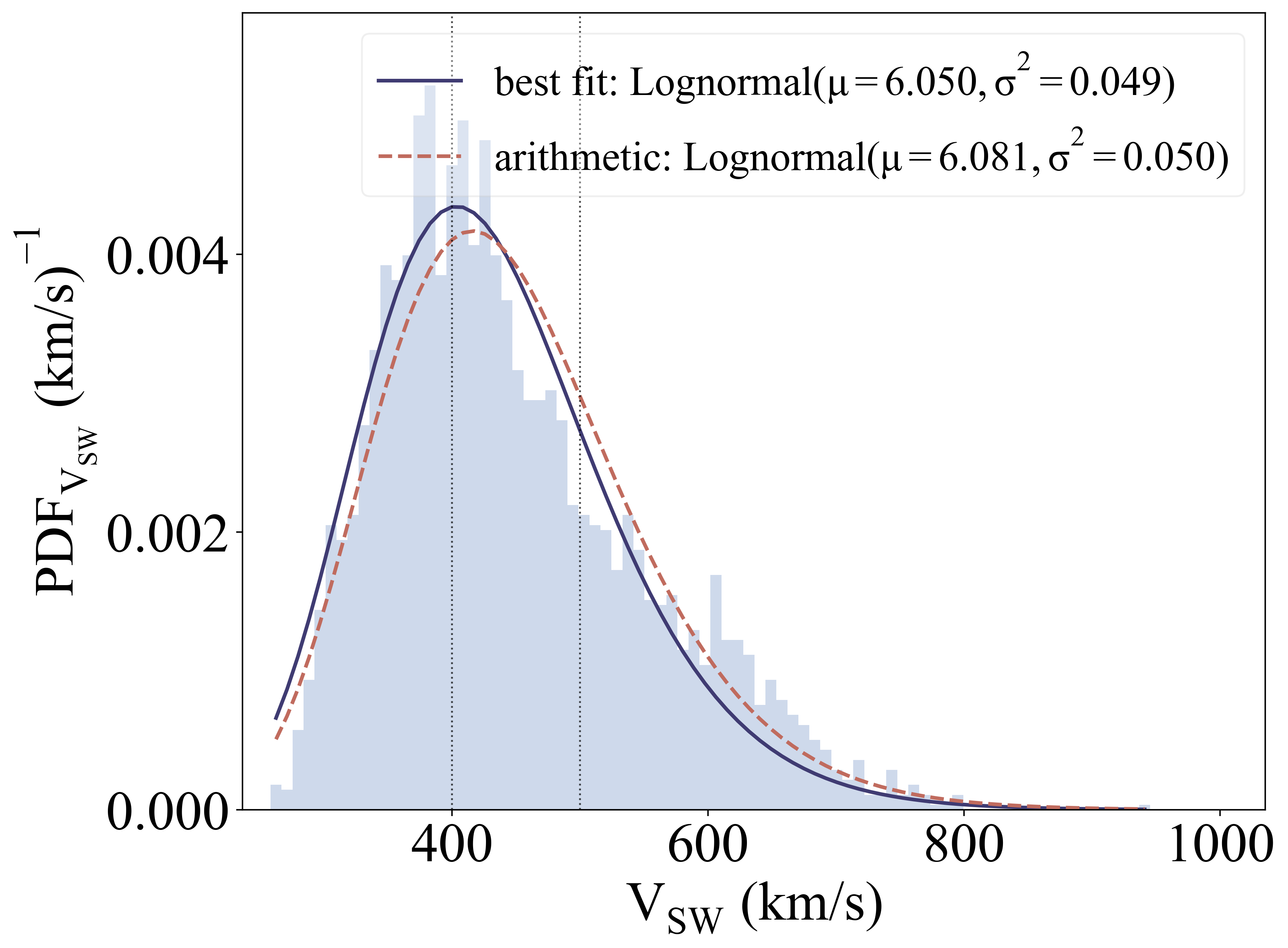}
     \caption{Shaded histogram shows the distribution of solar wind speeds. Dotted vertical lines represent the boundaries of the solar wind speed channels. Solid curve shows the best-fit lognormal distribution. Dashed curve shows the lognormal distribution derived from the arithmetic mean and variance of the wind speed samples. Parameters of both lognormal distributions are listed in the legend.}
\label{fig:Vdist}
\end{figure}

The separation of data intervals into three wind speed classes is a central part of this analysis. To ensure a reasonable level of statistical validity in each class, we base the partitioning by speed on an understanding of its probability distribution, as shown in 
Fig.~\ref{fig:Vdist}. 
The boundaries for slow, medium, and fast wind channels are represented by dotted vertical lines to show an almost equal number of counts in each channel.

\section{Statistics of solar wind density at 1 au}
\label{sec:app2}

\begin{figure}[!htbp]
\centering
    \includegraphics[angle=0,width=0.95\columnwidth]{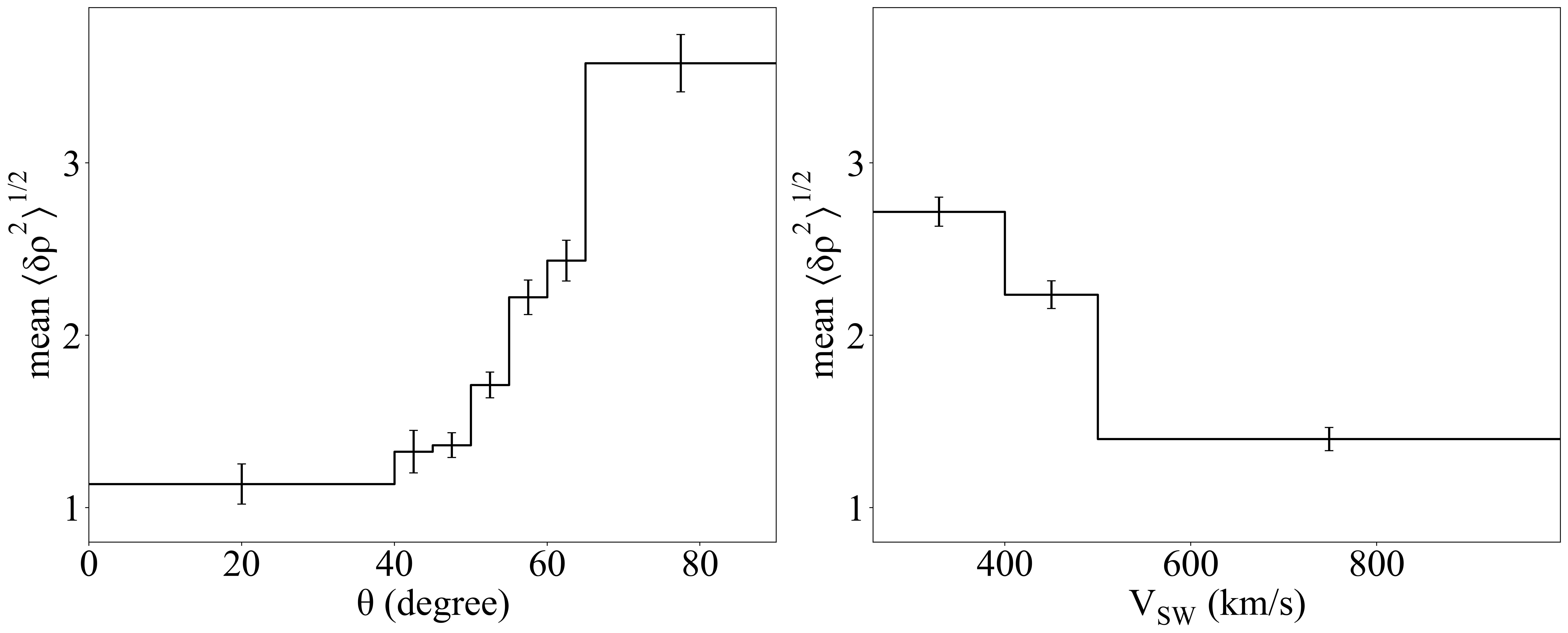}
     \caption{Density standard deviation averaged over each $\theta$ channel (left panel) and each $V_\mathrm{SW}$ channel (right panel). Error bars represent standard errors. Here $\delta \rho(t)  \equiv \rho (t) - \langle \rho(t)\rangle$, where \(\langle \cdots \rangle\) refers to an average over an individual 24-hr dataset, and ``mean'' refers to averaging the coefficient of variations over all 24-hour datasets that lie within a \(\theta\) or \(V_\mathrm{SW}\) channel.}
\label{fig:nStd}
\end{figure}

Here, we provide further statistics on our density samples. The standard deviation, S.D., measures the fluctuation amplitude, as opposed to the relative fluctuation amplitude shown in Fig.~\ref{fig:var_dist}. Fig.~\ref{fig:nStd} shows the average S.D. of the density samples with respect to $\theta$ and $V_\mathrm{SW}$, with error bars representing standard errors. The S.D. values range from $0.6$ to $3.4$ cm$^{-3}$ across the 24-hour datasets, and increase with $\theta$ and decrease with $V_\mathrm{SW}$. By comparing Fig.~\ref{fig:var_dist} and ~\ref{fig:nStd}, we note that the trend of the density fluctuation amplitude increasing with the angle between the magnetic and velocity fields is not compensated by the increasing mean density. Whereas the tendency towards higher fluctuation amplitude in slower winds is predominantly ascribed to a corresponding increase in mean density.

\begin{figure}[!htbp]
\centering
    \includegraphics[angle=0,width=0.6\columnwidth]{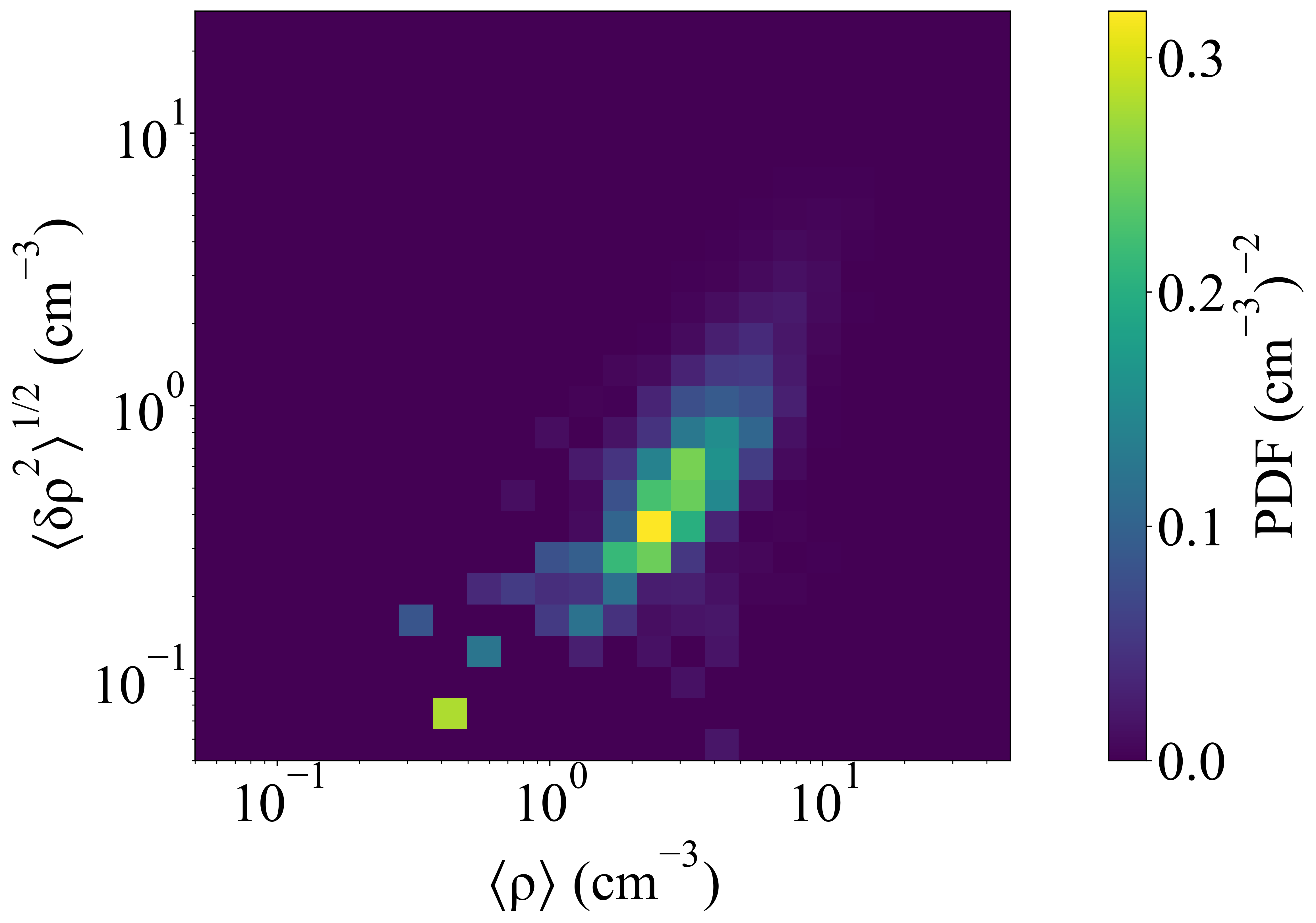}
     \caption{2D distribution of density standard deviation and mean. Data from all $\theta$ and $V_\mathrm{SW}$ channels are combined.}
\label{fig:stdmean}
\end{figure}

To study the relationship between mean density and density fluctuation in the solar wind at 1 au, we plot in Fig.~\ref{fig:stdmean} the joint probability density of the two variables. As expected, we observe larger mean densities corresponding to greater fluctuations.

\clearpage
%\bibliography{refs,ag,hl,mp,qz,refs_WHM, new_reference}

\begin{thebibliography}{}
\expandafter\ifx\csname natexlab\endcsname\relax\def\natexlab#1{#1}\fi
\providecommand{\url}[1]{\href{#1}{#1}}
\providecommand{\dodoi}[1]{doi:~\href{http://doi.org/#1}{\nolinkurl{#1}}}
\providecommand{\doeprint}[1]{\href{http://ascl.net/#1}{\nolinkurl{http://ascl.net/#1}}}
\providecommand{\doarXiv}[1]{\href{https://arxiv.org/abs/#1}{\nolinkurl{https://arxiv.org/abs/#1}}}

\bibitem[{{Adhikari} {et~al.}(2017){Adhikari}, {Zank}, {Hunana}, {Shiota},
  {Bruno}, {Hu}, \& {Telloni}}]{AdhikariEA17}
{Adhikari}, L., {Zank}, G.~P., {Hunana}, P., {et~al.} 2017, \apj, 841, 85,
  \dodoi{10.3847/1538-4357/aa6f5d}

\bibitem[{{Armstrong} {et~al.}(1990){Armstrong}, {Coles}, {Kojima}, \&
  {Rickett}}]{ArmstrongEA90}
{Armstrong}, J.~W., {Coles}, W.~A., {Kojima}, M., \& {Rickett}, B.~J. 1990,
  \apj, 358, 685, \dodoi{10.1086/169022}

\bibitem[{{Balmaceda} {et~al.}(2009){Balmaceda}, {Solanki}, {Krivova}, \&
  {Foster}}]{Balmaceda2009JGRA}
{Balmaceda}, L.~A., {Solanki}, S.~K., {Krivova}, N.~A., \& {Foster}, S. 2009,
  Journal of Geophysical Research (Space Physics), 114, A07104,
  \dodoi{10.1029/2009JA014299}

\bibitem[{{Barnes}(1981)}]{Barnes81}
{Barnes}, A. 1981, \jgr, 86, 7498, \dodoi{10.1029/JA086iA09p07498}

\bibitem[{{Barnes} \& {Hollweg}(1974)}]{BarnesHollweg74}
{Barnes}, A., \& {Hollweg}, J.~V. 1974, \jgr, 79, 2302,
  \dodoi{10.1029/JA079i016p02302}

\bibitem[{Batchelor(1970)}]{BatchelorTHT}
Batchelor, G.~K. 1970, The Theory of Homogeneous Turbulence (Cambridge, UK:
  Cambridge University Press)

\bibitem[{{Bellamy} {et~al.}(2005){Bellamy}, {Cairns}, \&
  {Smith}}]{BellamyEA05}
{Bellamy}, B.~R., {Cairns}, I.~H., \& {Smith}, C.~W. 2005, J.\ Geophys.\ Res.,
  110, A10104, \dodoi{10.1029/2004JA010952}

\bibitem[{{Bieber} {et~al.}(1994){Bieber}, {Matthaeus}, {Smith}, {Wanner},
  {Kallenrode}, \& {Wibberenz}}]{BieberEA94}
{Bieber}, J.~W., {Matthaeus}, W.~H., {Smith}, C.~W., {et~al.} 1994, \apj, 420,
  294, \dodoi{10.1086/173559}

\bibitem[{{Bieber} {et~al.}(1996){Bieber}, {Wanner}, \&
  {Matthaeus}}]{BieberEA96}
{Bieber}, J.~W., {Wanner}, W., \& {Matthaeus}, W.~H. 1996, \jgr, 101, 2511,
  \dodoi{10.1029/95JA02588}

\bibitem[{Blackman \& Tukey(1958)}]{BlackmanTukey}
Blackman, R.~B., \& Tukey, J.~W. 1958, The Measurement of Power Spectra (Dover)

\bibitem[{{Borovsky}(2012)}]{Borovsky12-spectra}
{Borovsky}, J.~E. 2012, Journal of Geophysical Research (Space Physics), 117,
  A05104, \dodoi{10.1029/2011JA017499}

\bibitem[{{Bruno} \& {Carbone}(2013)}]{BrunoCarboneLRSP13}
{Bruno}, R., \& {Carbone}, V. 2013, Living Reviews in Solar Physics, 10, 2,
  \dodoi{10.12942/lrsp-2013-2}

\bibitem[{{Cane} \& {Richardson}(2003)}]{CaneRichardson03}
{Cane}, H.~V., \& {Richardson}, I.~G. 2003, Journal of Geophysical Research
  (Space Physics), 108, 1156, \dodoi{10.1029/2002JA009817}

\bibitem[{{Celnikier} {et~al.}(1987){Celnikier}, {Muschietti}, \&
  {Goldman}}]{CelnikierEA87}
{Celnikier}, L.~M., {Muschietti}, L., \& {Goldman}, M.~V. 1987, \aap, 181, 138

\bibitem[{{Chandran}(2005)}]{Chandran05}
{Chandran}, B. D.~G. 2005, \prl, 95, 265004,
  \dodoi{10.1103/PhysRevLett.95.265004}

\bibitem[{{Chandran} \& {Backer}(2002)}]{Chandran02-radio}
{Chandran}, B.~D.~G., \& {Backer}, D.~C. 2002, \apj, 576, 176,
  \dodoi{10.1086/340792}

\bibitem[{{Chen} {et~al.}(2012){Chen}, {Mallet}, {Schekochihin}, {Horbury},
  {Wicks}, \& {Bale}}]{chen2012ApJ_density_aniso}
{Chen}, C.~H.~K., {Mallet}, A., {Schekochihin}, A.~A., {et~al.} 2012, \apj,
  758, 120, \dodoi{10.1088/0004-637X/758/2/120}

\bibitem[{{Chen} {et~al.}(2011){Chen}, {Mallet}, {Yousef}, {Schekochihin}, \&
  {Horbury}}]{ChenEA11-SWaniso}
{Chen}, C.~H.~K., {Mallet}, A., {Yousef}, T.~A., {Schekochihin}, A.~A., \&
  {Horbury}, T.~S. 2011, \mnras, 415, 3219,
  \dodoi{10.1111/j.1365-2966.2011.18933.x}

\bibitem[{{Cho} \& {Lazarian}(2002)}]{ChoLazarian02-prl}
{Cho}, J., \& {Lazarian}, A. 2002, \prl, 88, 245001,
  \dodoi{10.1103/PhysRevLett.88.245001}

\bibitem[{{Cho} {et~al.}(2002){Cho}, {Lazarian}, \& {Vishniac}}]{ChoEA02a}
{Cho}, J., {Lazarian}, A., \& {Vishniac}, E.~T. 2002, \apj, 564, 291,
  \dodoi{10.1086/324186}

\bibitem[{{Coles} \& {Harmon}(1989)}]{ColesHarmon89}
{Coles}, W.~A., \& {Harmon}, J.~K. 1989, \apj, 337, 1023,
  \dodoi{10.1086/167173}

\bibitem[{{Cuesta} {et~al.}(2022){Cuesta}, {Chhiber}, {Roy}, {Goodwill},
  {Pecora}, {Jarosik}, {Matthaeus}, {Parashar}, \&
  {Bandyopadhyay}}]{cuesta2022isotropization}
{Cuesta}, M.~E., {Chhiber}, R., {Roy}, S., {et~al.} 2022, \apjl, 932, L11,
  \dodoi{10.3847/2041-8213/ac73fd}

\bibitem[{{Dasso} {et~al.}(2005){Dasso}, {Milano}, {Matthaeus}, \&
  {Smith}}]{DassoEA05}
{Dasso}, S., {Milano}, L.~J., {Matthaeus}, W.~H., \& {Smith}, C.~W. 2005,
  \apjl, 635, L181, \dodoi{10.1086/499559}

\bibitem[{{Deforest} {et~al.}(2022){Deforest}, {Killough}, {Gibson}, {Henry},
  {Case}, {Beasley}, {Laurent}, {Colaninno}, {Waltham}, \& {Punch Science
  Team}}]{deforest2022PUNCH}
{Deforest}, C., {Killough}, R., {Gibson}, S., {et~al.} 2022, in 2022 IEEE
  Aerospace Conference, 1--11, \dodoi{10.1109/AERO53065.2022.9843340}

\bibitem[{{DeForest} {et~al.}(2016){DeForest}, {Matthaeus}, {Viall}, \&
  {Cranmer}}]{deforest2016fading}
{DeForest}, C.~E., {Matthaeus}, W.~H., {Viall}, N.~M., \& {Cranmer}, S.~R.
  2016, \apj, 828, 66, \dodoi{10.3847/0004-637X/828/2/66}

\bibitem[{{Du} {et~al.}(2023){Du}, {Li}, {Gan}, \& {Fu}}]{Du2023ApJ}
{Du}, S., {Li}, H., {Gan}, Z., \& {Fu}, X. 2023, \apj, 946, 74,
  \dodoi{10.3847/1538-4357/acc10b}

\bibitem[{{Frisch}(1995)}]{frisch1995turbulence}
{Frisch}, U. 1995, {Turbulence: the legacy of AN Kolmogorov} (Cambridge
  university press)

\bibitem[{{Gan} {et~al.}(2022){Gan}, {Li}, {Fu}, \& {Du}}]{GanEA22}
{Gan}, Z., {Li}, H., {Fu}, X., \& {Du}, S. 2022, \apj, 926, 222,
  \dodoi{10.3847/1538-4357/ac4d9d}

\bibitem[{{Germano}(1992)}]{Germano92}
{Germano}, M. 1992, Journal of Fluid Mechanics, 238, 325,
  \dodoi{10.1017/S0022112092001733}

\bibitem[{{Goldreich} \& {Sridhar}(1995)}]{GoldreichSridhar95}
{Goldreich}, P., \& {Sridhar}, S. 1995, \apj, 438, 763, \dodoi{10.1086/175121}

\bibitem[{{Hamilton} {et~al.}(2008){Hamilton}, {Smith}, {Vasquez}, \&
  {Leamon}}]{HamiltonEA08}
{Hamilton}, K., {Smith}, C.~W., {Vasquez}, B.~J., \& {Leamon}, R.~J. 2008,
  Journal of Geophysical Research (Space Physics), 113, A01106,
  \dodoi{10.1029/2007JA012559}

\bibitem[{{Horbury} {et~al.}(2012){Horbury}, {Wicks}, \& {Chen}}]{HorburyEA12}
{Horbury}, T.~S., {Wicks}, R.~T., \& {Chen}, C.~H.~K. 2012, \ssr, 172, 325,
  \dodoi{10.1007/s11214-011-9821-9}

\bibitem[{{Kellogg} \& {Horbury}(2005)}]{KelloggHorbury05}
{Kellogg}, P.~J., \& {Horbury}, T.~S. 2005, Annales Geophysicae, 23, 3765,
  \dodoi{10.5194/angeo-23-3765-2005}

\bibitem[{{Klein} \& {Burlaga}(1982)}]{KleinBurlaga82}
{Klein}, L.~W., \& {Burlaga}, L.~F. 1982, \jgr, 87, 613,
  \dodoi{10.1029/JA087iA02p00613}

\bibitem[{{Kontar} {et~al.}(2023){Kontar}, {Emslie}, {Clarkson}, {Chen},
  {Chrysaphi}, {Azzollini}, {Jeffrey}, \& {Gordovskyy}}]{kontar2023ApJ}
{Kontar}, E.~P., {Emslie}, A.~G., {Clarkson}, D.~L., {et~al.} 2023, \apj, 956,
  112, \dodoi{10.3847/1538-4357/acf6c1}

\bibitem[{{Malaspina} {et~al.}(2010){Malaspina}, {Kellogg}, {Bale}, \&
  {Ergun}}]{MalaspinaEA10}
{Malaspina}, D.~M., {Kellogg}, P.~J., {Bale}, S.~D., \& {Ergun}, R.~E. 2010,
  \apj, 711, 322, \dodoi{10.1088/0004-637X/711/1/322}

\bibitem[{{Matthaeus} {et~al.}(1996){Matthaeus}, {Ghosh}, {Oughton}, \&
  {Roberts}}]{MattEA96-var}
{Matthaeus}, W.~H., {Ghosh}, S., {Oughton}, S., \& {Roberts}, D.~A. 1996, \jgr,
  101, 7619, \dodoi{10.1029/95JA03830}

\bibitem[{{Matthaeus} \& {Goldstein}(1982)}]{matthaeus1982measurement}
{Matthaeus}, W.~H., \& {Goldstein}, M.~L. 1982, \jgr, 87, 6011,
  \dodoi{10.1029/JA087iA08p06011}

\bibitem[{{Matthaeus} {et~al.}(1990){Matthaeus}, {Goldstein}, \&
  {Roberts}}]{MattEA90}
{Matthaeus}, W.~H., {Goldstein}, M.~L., \& {Roberts}, D.~A. 1990, \jgr, 95,
  20673, \dodoi{10.1029/JA095iA12p20673}

\bibitem[{{Matthaeus} {et~al.}(2012){Matthaeus}, {Servidio}, {Dmitruk},
  {Carbone}, {Oughton}, {Wan}, \& {Osman}}]{MatthaeusEA12}
{Matthaeus}, W.~H., {Servidio}, S., {Dmitruk}, P., {et~al.} 2012, \apj, 750,
  103, \dodoi{10.1088/0004-637X/750/2/103}

\bibitem[{{McComas} {et~al.}(1998){McComas}, {Bame}, {Barker}, {Feldman},
  {Phillips}, {Riley}, \& {Griffee}}]{McComas98-swepam}
{McComas}, D.~J., {Bame}, S.~J., {Barker}, P., {et~al.} 1998, \ssr, 86, 563,
  \dodoi{10.1023/A:1005040232597}

\bibitem[{{McComas} {et~al.}(2000){McComas}, {Barraclough}, {Funsten},
  {Gosling}, {Santiago-Mu{\~n}oz}, {Skoug}, {Goldstein}, {Neugebauer}, {Riley},
  \& {Balogh}}]{McComasEA00}
{McComas}, D.~J., {Barraclough}, B.~L., {Funsten}, H.~O., {et~al.} 2000, \jgr,
  105, 10419, \dodoi{10.1029/1999JA000383}

\bibitem[{{Narita} {et~al.}(2010){Narita}, {Glassmeier}, {Sahraoui}, \&
  {Goldstein}}]{NaritaEA10-EbAniso}
{Narita}, Y., {Glassmeier}, K.~H., {Sahraoui}, F., \& {Goldstein}, M.~L. 2010,
  \prl, 104, 171101, \dodoi{10.1103/PhysRevLett.104.171101}

\bibitem[{{Oughton} \& {Matthaeus}(2020)}]{OughtonMatthaeus20}
{Oughton}, S., \& {Matthaeus}, W.~H. 2020, \apj, 897, 37,
  \dodoi{10.3847/1538-4357/ab8f2a}

\bibitem[{{Oughton} {et~al.}(2015){Oughton}, {Matthaeus}, {Wan}, \&
  {Osman}}]{OughtonEA15}
{Oughton}, S., {Matthaeus}, W.~H., {Wan}, M., \& {Osman}, K.~T. 2015,
  Philosophical Transactions of the Royal Society of London Series A, 373,
  20140152, \dodoi{10.1098/rsta.2014.0152}

\bibitem[{{Oughton} {et~al.}(1994){Oughton}, {Priest}, \&
  {Matthaeus}}]{OughtonEA94}
{Oughton}, S., {Priest}, E.~R., \& {Matthaeus}, W.~H. 1994, Journal of Fluid
  Mechanics, 280, 95, \dodoi{10.1017/S0022112094002867}

\bibitem[{Panchev(1971)}]{Panchev}
Panchev, S. 1971, Random Functions and Turbulence (New York: Pergammon Press)

\bibitem[{Richardson \& Cane(2024)}]{RichardsonCane24}
Richardson, I., \& Cane, H. 2024, {Near-Earth Interplanetary Coronal Mass
  Ejections Since January 1996}, V2,  Harvard Dataverse,
  \dodoi{10.7910/DVN/C2MHTH}

\bibitem[{{Robinson} \& {Rusbridge}(1971)}]{RobinsonRusbridge71}
{Robinson}, D.~C., \& {Rusbridge}, M.~G. 1971, Physics of Fluids, 14, 2499,
  \dodoi{10.1063/1.1693359}

\bibitem[{{Roy} {et~al.}(2021){Roy}, {Chhiber}, {Dasso}, {Ruiz}, \&
  {Matthaeus}}]{RoyEA21}
{Roy}, S., {Chhiber}, R., {Dasso}, S., {Ruiz}, M.~E., \& {Matthaeus}, W.~H.
  2021, \apjl, 919, L27, \dodoi{10.3847/2041-8213/ac21d2}

\bibitem[{{Saur} \& {Bieber}(1999)}]{SaurBieber99}
{Saur}, J., \& {Bieber}, J.~W. 1999, \jgr, 104, 9975,
  \dodoi{10.1029/1998JA900077}

\bibitem[{{Shaikh} \& {Zank}(2010)}]{shaikh2010turbulent}
{Shaikh}, D., \& {Zank}, G.~P. 2010, \mnras, 402, 362,
  \dodoi{10.1111/j.1365-2966.2009.15881.x}

\bibitem[{{Shalchi}(2009)}]{Shalchi}
{Shalchi}, A. 2009, {Nonlinear Cosmic Ray Diffusion Theories}, Vol. 362,
  \dodoi{10.1007/978-3-642-00309-7}

\bibitem[{{Shalchi} {et~al.}(2010){Shalchi}, {Li}, \& {Zank}}]{Shalchi10}
{Shalchi}, A., {Li}, G., \& {Zank}, G.~P. 2010, \apss, 325, 99,
  \dodoi{10.1007/s10509-009-0168-6}

\bibitem[{{Shebalin} {et~al.}(1983){Shebalin}, {Matthaeus}, \&
  {Montgomery}}]{shebalin1983JPP}
{Shebalin}, J.~V., {Matthaeus}, W.~H., \& {Montgomery}, D. 1983, Journal of
  Plasma Physics, 29, 525, \dodoi{10.1017/S0022377800000933}

\bibitem[{{Smith} {et~al.}(1998){Smith}, {L'Heureux}, {Ness}, {Acu{\~n}a},
  {Burlaga}, \& {Scheifele}}]{Smith98-mag}
{Smith}, C.~W., {L'Heureux}, J., {Ness}, N.~F., {et~al.} 1998, \ssr, 86, 613,
  \dodoi{10.1023/A:1005092216668}

\bibitem[{{Taylor}(1938)}]{Taylor38}
{Taylor}, G.~I. 1938, Proceedings of the Royal Society of London Series A, 164,
  476, \dodoi{10.1098/rspa.1938.0032}

\bibitem[{{Usmanov} {et~al.}(2018){Usmanov}, {Matthaeus}, {Goldstein}, \&
  {Chhiber}}]{UsmanovEA18}
{Usmanov}, A.~V., {Matthaeus}, W.~H., {Goldstein}, M.~L., \& {Chhiber}, R.
  2018, \apj, 865, 25, \dodoi{10.3847/1538-4357/aad687}

\bibitem[{{V{\"o}lk} \& {Aplers}(1973)}]{VolkAlpers73}
{V{\"o}lk}, H.~J., \& {Aplers}, W. 1973, \apss, 20, 267,
  \dodoi{10.1007/BF00642204}

\bibitem[{{Weygand} {et~al.}(2011){Weygand}, {Matthaeus}, {Dasso}, \&
  {Kivelson}}]{WeygandEA11}
{Weygand}, J.~M., {Matthaeus}, W.~H., {Dasso}, S., \& {Kivelson}, M.~G. 2011,
  Journal of Geophysical Research (Space Physics), 116, A08102,
  \dodoi{10.1029/2011JA016621}

\bibitem[{{Zank} {et~al.}(2012){Zank}, {Jetha}, {Hu}, \&
  {Hunana}}]{ZankEA12-density}
{Zank}, G.~P., {Jetha}, N., {Hu}, Q., \& {Hunana}, P. 2012, \apj, 756, 21,
  \dodoi{10.1088/0004-637X/756/1/21}

\bibitem[{{Zank} {et~al.}(2004){Zank}, {Li}, {Florinski}, {Matthaeus}, {Webb},
  \& {Le Roux}}]{Zank04}
{Zank}, G.~P., {Li}, G., {Florinski}, V., {et~al.} 2004, Journal of Geophysical
  Research (Space Physics), 109, A04107, \dodoi{10.1029/2003JA010301}

\bibitem[{{Zank} {et~al.}(2023){Zank}, {Zhao}, {Adhikari}, {Nakanotani},
  {Pit{\v{n}}a}, {Telloni}, \& {Che}}]{ZankEA23-linear}
{Zank}, G.~P., {Zhao}, L.~L., {Adhikari}, L., {et~al.} 2023, \apjs, 268, 18,
  \dodoi{10.3847/1538-4365/acdf5d}
  
\bibitem[{{Zank} {et~al.}(2024){Zank}, {Zhao}, {Adhikari}, {Telloni},
  {Baruwal}, {Baruwal}, {Zhu}, {Nakanotani}, {Pitna}, {Kasper}, \&
  {Bale}}]{Zank24arXiv}
{Zank}, G.~P., {Zhao}, L., {Adhikari}, L., {et~al.} 2024, arXiv e-prints,
  arXiv:2403.14861, \dodoi{10.48550/arXiv.2403.14861}

\bibitem[{{Zhao} {et~al.}(2017){Zhao}, {Adhikari}, {Zank}, {Hu}, \&
  {Feng}}]{Zhao17}
{Zhao}, L.~L., {Adhikari}, L., {Zank}, G.~P., {Hu}, Q., \& {Feng}, X.~S. 2017,
  \apj, 849, 88, \dodoi{10.3847/1538-4357/aa932a}

\bibitem[{{Zhao} {et~al.}(2018){Zhao}, {Adhikari}, {Zank}, {Hu}, \&
  {Feng}}]{Zhao18}
---. 2018, \apj, 856, 94, \dodoi{10.3847/1538-4357/aab362}

\bibitem[{{Zhao} {et~al.}(2023){Zhao}, {Zank}, {Nakanotani}, \&
  {Adhikari}}]{Zhao23}
{Zhao}, L.~L., {Zank}, G.~P., {Nakanotani}, M., \& {Adhikari}, L. 2023, \apj,
  944, 98, \dodoi{10.3847/1538-4357/acb33b}

\end{thebibliography}

 \newcommand{\BIBand} {and} %...... how 'and' appears in authors
  \newcommand{\boldVol}[1] {\textbf{#1}} %......................
  \providecommand{\SortNoop}[1]{} %.......Use as {\SortNoop{Aaa}}
  \providecommand{\sortnoop}[1]{} %..............................
  \newcommand{\stereo} {\emph{{S}{T}{E}{R}{E}{O}}} %.................
  \newcommand{\au} {{A}{U}\ } %..................................
  \newcommand{\AU} {{A}{U}\ } %.................................
  \newcommand{\MHD} {{M}{H}{D}\ } %..............................
  \newcommand{\mhd} {{M}{H}{D}\ } %...............................
  \newcommand{\RMHD} {{R}{M}{H}{D}\ } %...........................
  \newcommand{\rmhd} {{R}{M}{H}{D}\ } %...........................
  \newcommand{\wkb} {{W}{K}{B}\ } %..............................
  \newcommand{\alfven} {{A}lfv{\'e}n\ } %...........................
  \newcommand{\alfvenic} {{A}lfv{\'e}nic\ } %.........................
  \newcommand{\Alfven} {{A}lfv{\'e}n\ } %...........................
  \newcommand{\Alfvenic} {{A}lfv{\'e}nic\ }

%\textbf{\small RC: I spoke to some people who are very interested in this result, we should get this submitted soon! It's of high relevance to astrophysics because density fluctuations play a bigger role there. The specific people I spoke to were interested in the scattering of radiation from distant objects when that radiation passes through a field of density turbulence and magnetic field. The point source image gets stretched in a direction depending on the density fluctuation anisotropy xx}

%\listofchanges
\end{document}